\documentclass[twocolumn,english,prl,showpacs,aps,letterpaper,superscriptaddress]{revtex4-1}

\usepackage[T1]{fontenc}
\usepackage[latin9]{inputenc}
\usepackage{amsmath}
\usepackage{amssymb}
\usepackage{graphicx}
\usepackage{hyperref}
\usepackage{natbib}
\usepackage{times}
\usepackage{helvet}

\makeatletter
 
 \@ifundefined{textcolor}{}
 {%
   \definecolor{BLACK}{gray}{0}
   \definecolor{WHITE}{gray}{1}
   \definecolor{RED}{rgb}{1,0,0}
   \definecolor{GREEN}{rgb}{0,1,0}
   \definecolor{BLUE}{rgb}{0,0,1}
   \definecolor{CYAN}{cmyk}{1,0,0,0}
   \definecolor{MAGENTA}{cmyk}{0,1,0,0}
   \definecolor{YELLOW}{cmyk}{0,0,1,0}
 }

\@ifundefined{textcolor}{}{%
 \definecolor{BLACK}{gray}{0}
 \definecolor{WHITE}{gray}{1}
 \definecolor{RED}{rgb}{1,0,0}
 \definecolor{GREEN}{rgb}{0,1,0}
 \definecolor{BLUE}{rgb}{0,0,1}
 \definecolor{CYAN}{cmyk}{1,0,0,0}
 \definecolor{MAGENTA}{cmyk}{0,1,0,0}
 \definecolor{YELLOW}{cmyk}{0,0,1,0}
 }

\@ifundefined{definecolor}{\@ifundefined{definecolor}
 {\@ifundefined{definecolor}
 {\@ifundefined{definecolor}
 {\@ifundefined{definecolor}
 {\@ifundefined{definecolor}
 {\usepackage{color}}{}
}{}
}{}
}{}
}{}
}{}
\makeatother

\begin{document}

\title{Robust spin squeezing via photon-mediated interactions on an optical clock transition}

\author{R.~J.~Lewis-Swan}
\affiliation{JILA, NIST, and Dept. of Physics, University of Colorado, 440 UCB, Boulder, CO  80309, USA}
\affiliation{Center for Theory of Quantum Matter, University of Colorado, Boulder, CO 80309, USA}
\author{M.~A.~Norcia}
\affiliation{JILA, NIST, and Dept. of Physics, University of Colorado, 440 UCB, Boulder, CO  80309, USA}
\author{J.~R.~K.~Cline}
\affiliation{JILA, NIST, and Dept. of Physics, University of Colorado, 440 UCB, Boulder, CO  80309, USA}
\author{J.~K.~Thompson}
\affiliation{JILA, NIST, and Dept. of Physics, University of Colorado, 440 UCB, Boulder, CO  80309, USA}
\author{A.~M.~Rey}
\affiliation{JILA, NIST, and Dept. of Physics, University of Colorado, 440 UCB, Boulder, CO  80309, USA}
\affiliation{Center for Theory of Quantum Matter, University of Colorado, Boulder, CO 80309, USA}

\date{\today }
\begin{abstract}


Cavity-QED is a promising avenue for the deterministic generation of entangled and spin-squeezed states for quantum metrology. One archetypal scheme generates squeezing via collective one-axis
twisting interactions. However, we show that in implementations using optical transitions in long-lived atoms the achievable squeezing is fundamentally limited by collectively enhanced emission into the
cavity mode which is generated in parallel with the cavity-mediated spin-spin interactions. We propose an alternative scheme which generates a squeezed state that is protected from collective emission,
and investigate its sensitivity to realistic sources of experimental noise and imperfections.


\end{abstract}

\maketitle

\noindent{\it Introduction:} Atomic clocks operated with long-lived optically excited states in large ensembles of alkaline-earth atoms have led to unprecendent advances in frequency and time-standards
\cite{Bloom2014,Campbell2017,Ludlow2015,Grebing2016,Lodewyck2016,Takano2016}.
This has been achieved  both by taking advantage of the superior precision afforded by operating at optical rather than microwave frequencies, 
and by utilizing large numbers of atoms $N$ to quickly average down quantum projection noise.
However, these clocks are reaching a point where improvements in sensing capabilities based on individual particle control have limited return due to physical and practical constraints, such as difficulty
increasing the number of participating atoms due to collisional shifts \cite{Martin2013}. This presents a clear need for a new paradigm of sensors that utilize many-particle quantum correlations \cite{Wineland1992}, dramatically reducing quantum
noise and breaking through the standard quantum limit (SQL) on phase-sensitivity, $\delta\phi \sim 1/\sqrt{N}$~rad. However, quantum correlations are difficult to create and intrinsically fragile to decoherence, and
therefore the design and implementation of robust methods for entanglement generation is an important current challenge for quantum-enhanced sensors, and in particular,
the next-generation of atomic clocks.

%
%

A canonical example of useful entangled states for metrology are squeezed states \cite{Walls1983,Ueda_SpinSqueezing_1993}, which feature a reduction of the quantum projection noise along a particular
quadrature. In atomic ensembles, spin-squeezed states have successfully been generated in proof-of-principle systems that operate on microwave-frequency transitions by projective measurement and feedback
protocols \cite{Appel2009,SchleiherSmith2010,Chen2011,Bohnet2014,Cox2016,Hosten2016}, with state-of-the-art schemes reaching $\sim 18$~dB below SQL \cite{Cox2016,Hosten2016}. Deterministic production of
spin-squeezed states generated by one-axis twisting (OAT) schemes has also been demonstrated on microwave transitions \cite{Leroux2010,HostenPhaseMagnification2016,Gross2010}. However, the best reported squeezing remains
limited at $8$~dB below SQL \cite{HostenPhaseMagnification2016}. It is then desirable to understand how entanglement generated by unitary dynamics can be significantly improved, particularly protocols 
applicable to optical transitions used in current state-of-the-art atomic clocks.

In this vein, recent work has highlighted the possibility of using photon-mediated spin-exchange interactions to engineer OAT in an undriven
optical cavity \cite{Norcia2017,Vuletic2017}. The scheme is relevant to the dynamical generation of spin-squeezed states directly on the narrow linewidth optical clock transition.
However, the achievable squeezing is severely limited by intrinsic dissipative noise arising due to superradiance: the collective emission and leakage of photons from the cavity.

In this manuscript, we propose to overcome this problem by generating squeezing from an unorthodox initial state composed of a pair of spin ensembles with zero mean total spin-projection. In this protocol,
which we refer to as `two-spin squeezing' (TSS), the squeezing is generated in an almost orthogonal quadrature to the noise arising from superradiance. Our theoretical calculations demonstrate this leads to a
reduced sensitivity to collective emission and consequently the TSS scheme out-performs the conventional OAT protocol. We also examine the robustness of TSS in the presence of typical sources
of single-particle decoherence.

\begin{figure}
 \includegraphics[width=8cm]{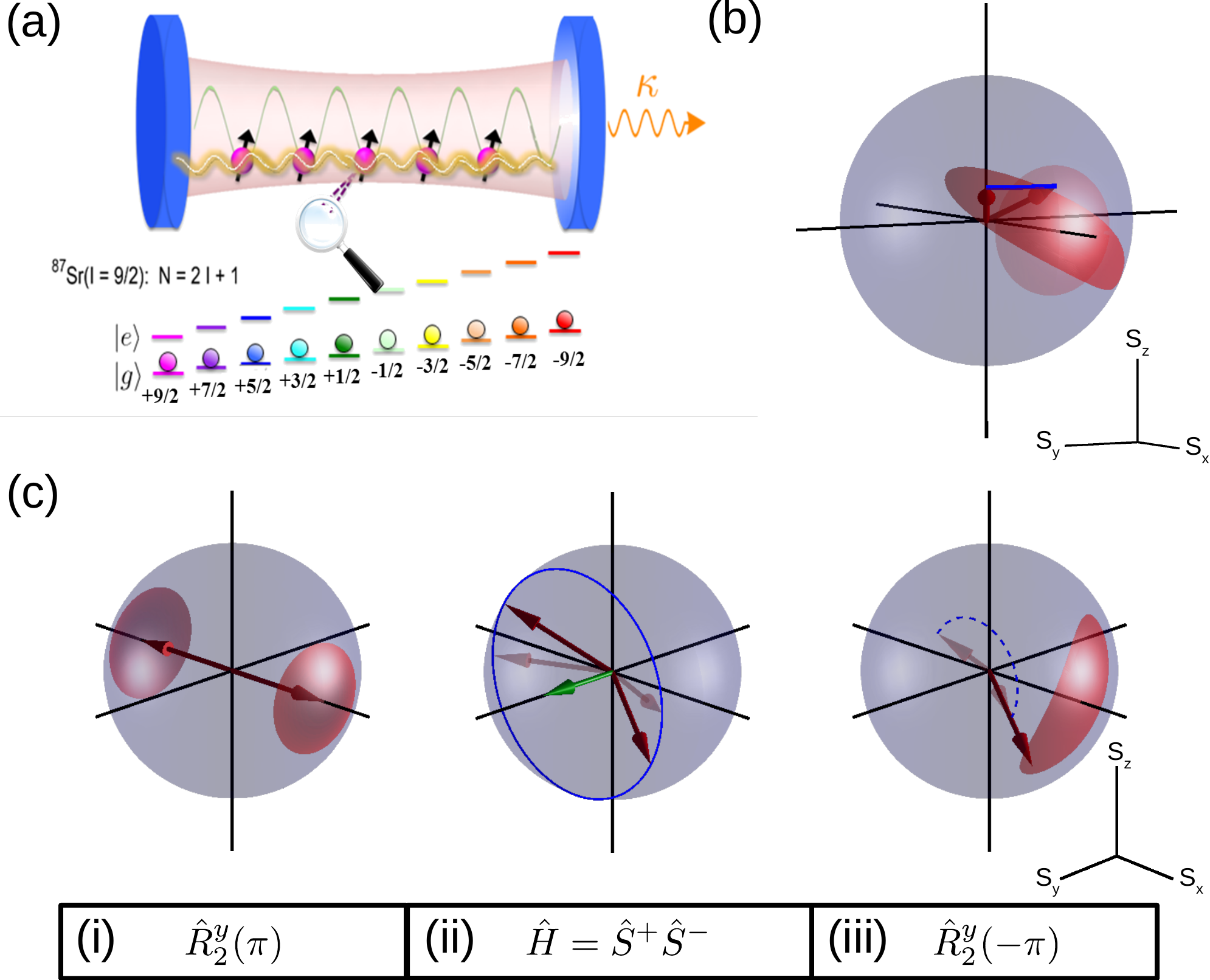}
 \caption{(a) Proposed experimental system, consisting of an ensemble of $N$ atoms trapped in a standing-wave lattice potential and optically coupled to the field within an optical cavity.
 The multiple $m_F$ spin-projections of the $^1$S$_0$ to $^3$P$_0$ transition of $^{87}$Sr \cite{Norcia2017} allow the realization of multiple independently controllable collective spins.
 (b) Cavity-mediated one-axis twisting. Quantum fluctuations along $S_z$ (light distribution) lead to a precession of the Bloch vector $\vec{S}$ (red, trajectory indicated by blue) about the $S_z$-axis, generating
 squeezing of the noise distribution (dark distribution).
 (c) Two-spin squeezing. (i) and (ii) Common-mode fluctuations along $S_y$ of a pair of back-to-back collective spins (faded red) generates a weak coherence along $S_y$ (green), which both drives and defines the precession 
 axis of the Bloch vectors $\vec{S}_{1,2}$ (solid red, blue circle guides the eye for the precesion). (iii) Subsequent rotation about $S_y$ of one of the spins (indicated by dashed blue arc)
 maps this precession into a squeezing of the collective distribution.}
 \label{fig:Schematic}
\end{figure}

\noindent{\it Model and definitions:} We consider a system of $N$ atoms trapped in a standing-wave optical lattice which is supported by an optical cavity \cite{Norcia2017}, 
illustrated in Fig.~\ref{fig:Schematic}(a).
The cavity field couples the ground and excited clock states of the atom with single-photon Rabi frequency $2g$. We assume the atom-light coupling is spatially uniform throughout the cavity,
such that we can describe the atomic ensemble using collective spin operators $\hat{S}^{x,y,z}_{\alpha} \equiv \sum_{j} \hat{\sigma}^{x,y,z}_{j,\alpha}/2$ where $\hat{\sigma}^{x,y,z}_{j,\alpha}$ denote Pauli matrices. 
The summation of $j$ runs over the atomic ensemble and $\alpha$ indexes the possible internal degrees of freedom, e.g., hyperfine levels.

The narrow linewidth  of the clock transition,$\gamma$, relative to the cavity linewidth $\kappa\gg\gamma$, allows us to adiabatically  eliminate the cavity field such that the photons 
only mediate effective spin dynamics \cite{Norcia2017}. The reduced density matrix of the atomic spin then evolves according to the effective spin Hamiltonian (see SM)
\begin{equation}
\hat{H}_{\mathrm{eff}} = \hbar\sum_{\alpha,\beta} \chi_{\alpha,\beta} \hat{S}^+_{\alpha} \hat{S}^-_{\beta} , \label{eqn:spin_H}
\end{equation} 
and  the Lindblad jump operator $\hat{L}_{\Gamma} =  \sum_{\alpha} \sqrt{\Gamma_{\alpha}/2} \hat{S}^-_{\alpha}$, where
$\hat{S}^{\pm}_{\alpha} \equiv \hat{S}^x_{\alpha} + i\hat{S}^y_{\alpha}$  accounts for the collectively enhanced emission into the cavity. 
The relative strength of the elastic interactions $\chi_{\alpha,\beta} = 4g_{\alpha}g_{\beta}\Delta_c/(4\Delta_c^2 + \kappa^2)$ and dissipation
$\Gamma_{\alpha} = 4g_{\alpha}^2\kappa/(4\Delta_c^2 + \kappa^2)$ is controlled by the detuning $\Delta_c$ of the cavity from resonance with the atomic transition. 
Throughout the manuscript, we will now  set $\hbar = 1$

In terms of spin operators, squeezing is characterised by the parameter $\xi^2 \equiv N \mathrm{min}[\langle (\delta \hat{S}_{\psi} )^2 \rangle]/|\langle \vec{S} \rangle|^2$ \cite{Wineland1994},
where $\mathrm{min}[\langle (\delta \hat{S}_{\psi} )^2 \rangle]$ is the minimal quadrature variance of the state along a direction $\hat{n}_\psi$ perpendicular to 
$\langle \vec{S} \rangle$ ( i.e.  $\hat{n}_{\psi} \cdot  \langle \vec{S} \rangle =0$ and 
$\langle (\delta \hat{S}_{\psi} )^2 \rangle\equiv \langle (\hat{n}_{\psi} \cdot \hat{S}_{\psi} )^2 \rangle-\langle \hat{n}_{\psi} \cdot \hat{S}_{\psi}  \rangle ^2$). Squeezing $\xi^2 < 1$ indicates that the 
quantum noise of the state along one quadrature is reduced below the SQL, defined with respect to an isotropic coherent spin state.

\noindent{\it Spin-squeezing by one-axis twisting:} When a single internal level is populated, the effective Hamiltonian reduces to $\hat{H}_{\mathrm{eff}} = \chi \hat{S}^+\hat{S}^-$ which can be rewritten
as $\hat{H}_{\mathrm{eff}} \equiv \chi(\hat{S}^2 - \hat{S}_z^2 + \hat{S}_z)$. The term $\propto \hat{S}_z^2$ generates one-axis twisting \cite{Ueda_SpinSqueezing_1993}
whilst the last term generates a trivial single-particle rotation which can be neglected herein. The first term $\propto \hat{S}^2 \equiv \hat{S}_x^2 + \hat{S}_y^2 + \hat{S}_z^2$, commutes with the one-axis twisting,
but is responsible for opening a many-body gap between Dicke manifolds with different eigenvalues $S(S+1)$ ($S=0,...,N/2$) of $\hat{S}^2$ which can protect the collective dynamics from slow
single-particle decoherence \cite{Norcia2017}. Here, we assume the unitary dynamics is restricted to the $S=N/2$ manifold and thus consider $\hat{H}_{\mathrm{eff}}$ equivalent to the OAT 
Hamiltonian $\hat{H}_{\mathrm{OAT}} = \chi \hat{S}_z^2$.

The OAT spin squeezing can be understood in a semi-classical picture, illustrated in Fig.~\ref{fig:Schematic}~(b), in terms of the mean-field Hamiltonian 
$\hat{H}_{\mathrm{MF}} \equiv 2\chi \langle \hat{S}_z \rangle \hat{S}_z$, which generates rotations about the $S_z$-axis at a rate dependent on the atomic inversion.
Under $\hat{H}_{\mathrm{MF}}$  the isotropic noise distribution  of an initially prepared  spin coherent state along $x$, $|\Phi^{OAT}\rangle=|N/2\rangle_x$  with  $\hat{S}_x|N/2\rangle_x=N/2|N/2\rangle_x$,  
will shear into an anisotropic distribution with reduced noise along one quadrature and increased noise along the other. 
As the spin-spin interactions responsible for the OAT dynamics are mediated by a macroscopically populated cavity field, they are also accompanied by superradiant collective emission from
the cavity mode. 
A net leakage of photons from the cavity carries away information at the enhanced rate $\kappa \langle \hat{a}^\dagger \hat{a} \rangle\sim \kappa  N^2$, and correspondingly introduces
excess dissipative noise $\propto N^2\Gamma t$ to the $S_z$ quadrature, degrading the spin squeezing.

A perturbative treatment of both collective emission and the unitary dynamics leads to an expression for the time evolution of the squeezing (see Ref.~\cite{Vuletic2017} and SM),
\begin{equation}
 \xi^2_{\mathrm{OAT}} \approx \frac{1}{2N\beta} + \frac{2}{3}\beta^2 + \Gamma N t \label{eqn:OATpert_superrad}
\end{equation}
where  $\beta\equiv N\chi^2 t^2/2 \ll 1$. The term $\propto 1/\beta$ describes the spin squeezing, while the term $\propto \beta^2$ describes over-squeezing due to the curvature of 
the Bloch sphere that yields a non-Gaussian distribution. The last term $\Gamma N t$ describes the collectively enhanced dissipative noise added to the squeezed quadrature. 
For $\Gamma = 0$, the optimal achievable OAT squeezing is limited only by the curvature
of the Bloch sphere, which  is reached when $\beta^2 \sim 1/(N\beta)$ and for large atom number $N \gg 1$  scales as $\sim N^{-2/3}$ \cite{Ueda_SpinSqueezing_1993}.  
In contrast, superradiant emission, $\Gamma \neq 0$, limits the optimal squeezing when the added noise becomes comparable to $\Gamma N t \sim \frac{1}{N\beta}$. This noise typically becomes dominant at 
much earlier times than non-Gaussian effects. The impact of superradiance is more clear if one minimizes Eq.~(\ref{eqn:OATpert_superrad}) with respect to $t$, 
for fixed cavity parameter $\Gamma/\chi$. 
In this case, ignoring the term $\propto \beta^2$ in Eq.~(\ref{eqn:OATpert_superrad}) as negligible, the achievable squeezing is bounded by 
\begin{eqnarray}
 \left. \xi^2_{\mathrm{OAT}}\right\vert_{\Gamma,\chi} \approx \frac{3}{2^{2/3}} \left(\frac{\Gamma}{\chi}\right)^{2/3} , \label{eqn:OneSpinSq_FixGam}
\end{eqnarray} 
independent of atom number $N$.



\begin{figure}
  \includegraphics[width=8cm]{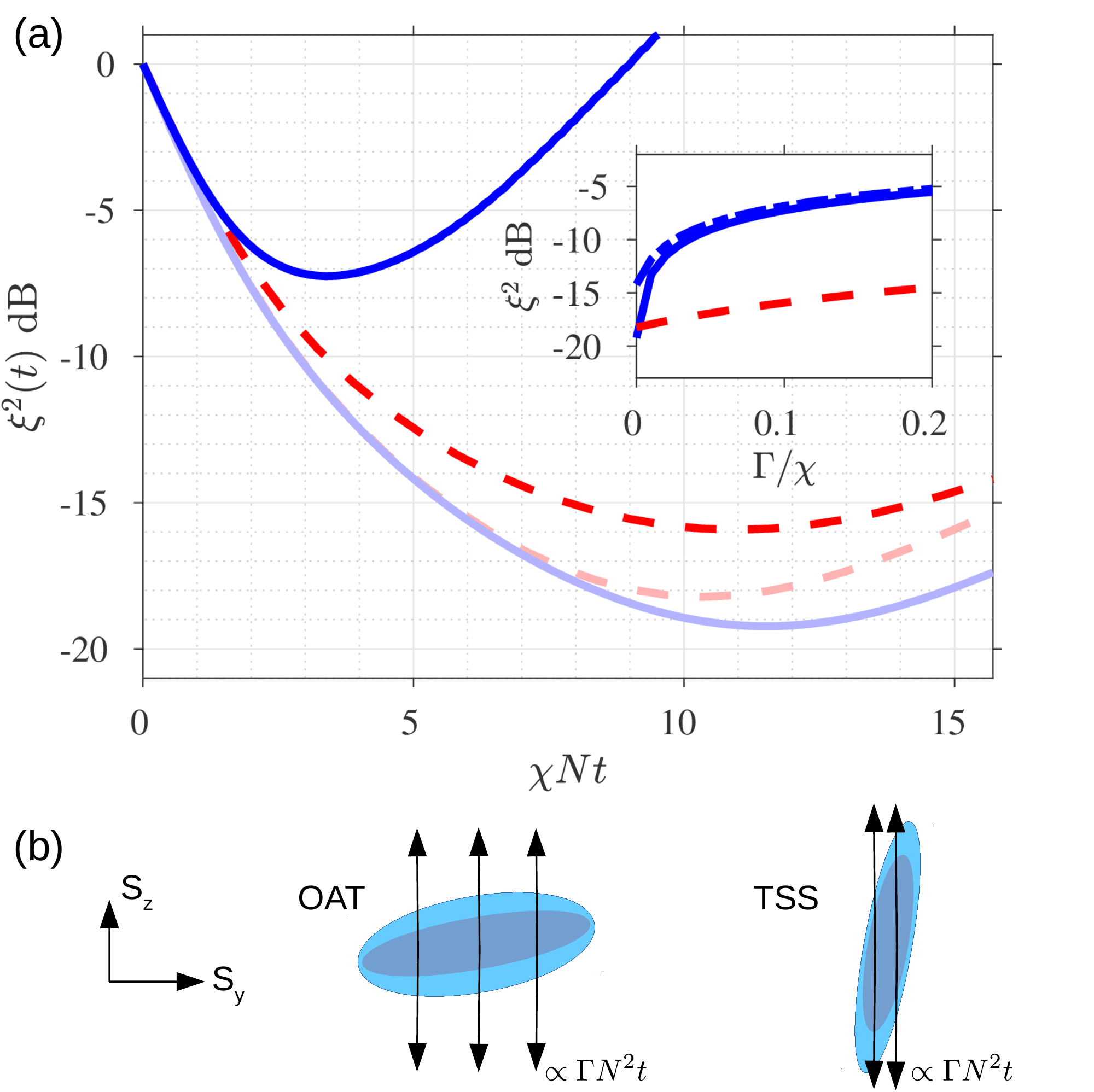}
 \caption{(a) Comparison of squeezing with OAT and TSS protocols for $N = 1000$. Faded lines are the ideal ($\Gamma = 0$) results for OAT (blue) and TSS (red dashed).
 Solid lines are $\Gamma = 0.1\chi$ results for OAT (blue) and TSS (red dashed). Best squeezing as a function of $\Gamma/\chi$ (solid blue -- OAT, dashed red -- TSS) is plotted in the inset. Dot-dashed blue line indicates
 squeezing obtained with OAT and $N=200$, illustrating the invariance with $N$.
 All calculations are numerical solution of the master equation as discussed in main text.
 (b) Dark region indicates ideal squeezed state, light region indicates distribution with added dissipative noise due to photon leakage from the cavity (arrows),
 which approximately aligns with the squeezed quadrature (OAT) or anti-squeezed quadrature (TSS).}
 \label{fig:SqFixGamma}
\end{figure}

\noindent{\it Two-spin squeezing (TSS):} We now consider the alternative scheme and demonstrate how 
initiating the dynamics from an unconventional state of zero mean coherence maps to an effective Hamiltonian which generates squeezing that is robust to collective emission.
Initially the atoms are separated in two different ensembles, denoted as $\alpha=1,2$, each composed of $N/2$ atoms, and the system is then prepared in an incoherent state composed of a 
pair of opposite (back-to-back) collective spins: $|\Phi^{TSS}\rangle=\vert N/4 \rangle_{x_1}\bigotimes \vert -N/4 \rangle_{x_2}$, 
each one in an stretched eigenstate of $\hat{S}^x_{j=1,2}$ but aligned  along opposing directions,  $\hat{S}^x_{j} \vert \pm N/4 \rangle_{x_j} \equiv \pm(N/4) \vert \pm N/4 \rangle_{x_j}$.
Such a state could be realized by utilizing the $m_F = \pm 9/2$ hyperfine  levels of $^{87}$Sr \cite{Norcia2017}. This initial state has zero mean coherence $\langle \hat{S}^+ \rangle = 0$ and 
$\langle \hat{S}^+ \hat{S}^-\rangle = N/2$, and as such the cavity occupation $\langle \hat{a}^{\dagger} \hat{a} \rangle \propto \langle \hat{S}^{+}\hat{S}^{-} \rangle \sim N$ is reduced by a factor of $N$ to 
the level of $N$ independent emitting atoms. For clarity, the total collective spin operators are $\hat{S}^{\alpha} \equiv \hat{S}^{\alpha}_1 + \hat{S}^{\alpha}_2$ for $\alpha=x,y,z$, 
where the subscript refers to the ensemble internal degree of freedom.

Even though the initial state has zero mean coherence it will nevertheless non-trivially evolve under the Hamiltonian 
$\hat{H}_{\mathrm{eff}} = \chi \hat{S}^+\hat{S}^- \equiv \chi (\hat{S}^+_1 + \hat{S}^+_2)(\hat{S}^-_1 + \hat{S}^-_2)$  [Eq. (1)]. 
This form of the Hamiltonian assumes  $\chi_{\alpha,\beta} \equiv \chi$,  satisfied for $m_F =  \pm 9/2$ or  any `symmetric' pair of hyperfine levels $\pm m_F$.
To reveal squeezing, after the dynamics we perform a local spin-flip rotation of the 2nd collective spin about $\hat{y}$. For short evolution under $\hat{H}_{\mathrm{eff}}$, which is what we consider 
in the following, the evolved state is approximately transferred back to the fully symmetric manifold $S=N/2$ (up to negligible corrections, see later discussion and SM)
For example, in the absence of any evolution under $\hat{H}_{\mathrm{eff}}$ the final state would be $|N/2\rangle_x$.

The overall protocol can be recast in terms of evolution under a Hamiltonian in a rotated reference-frame acting on an initially collective state ($S=N/2$) with all spins aligned together along $S_x$:
\begin{equation}
 \vert \psi(t) \rangle  = \hat{R}^y_{2}(-\pi) e^{-i\hat{H}_{\mathrm{eff}} t} \hat{R}^y_{2}(\pi) |N/2\rangle_x
 \equiv  e^{-i\hat{\tilde{H}} t} |N/2\rangle_x. \notag
\end{equation}
Here, $\hat{R}^y_{j}(\phi) = e^{-i\phi \hat{S}^y_j}$ is a collective spin rotation acting on the $j=1,2$ internal state and
$\hat{\tilde{H}} \equiv \chi\left[ (\hat{S}^x_1 - \hat{S}^x_2)^2 + (\hat{S}^y_1 + \hat{S}^y_2)^2 \right]$. The second term of $\hat{\tilde{H}}$ induces one-axis twisting about
the $\hat y$ axis, leading to an approximate azimuthally (`phase') squeezed state. The first term is more complex and could lead to degradation of squeezing correlations.
However, as the initial state in this rotated frame is an eigenstate of $\hat{S}^x_1$ and $\hat{S}^x_2$, then at the relevant short time-scale of squeezing 
this term can be ignored and the dynamics effectively remains in the $S=N/2$ manifold. Moreover, our scheme is not overly sensitive to this assumption and can tolerate
independent number fluctuations $\lesssim N^{1/3}$ in the prepared ensembles (see SM).

Physical intuition can be gained by a semi-classical description in the original frame of the back-to-back spins $|\Phi^{TSS}\rangle$,
as illustrated in Fig.~\ref{fig:Schematic} (c). For this initial state, the mean-field Hamiltonian corresponds to a precession of each of the individual spins about the $S_y$ projection of the
total collective spin, $\hat{H}_{\mathrm{MF}} \approx 2\chi \langle \hat{S}^y_1 + \hat{S}^y_2 \rangle \left( \hat{S}^y_1 + \hat{S}^y_2 \right)$.
Shearing is then induced by common-mode fluctuations of the initial
states $\sim \sqrt{N}$ along $S_y$ (i.~e., phase noise), which generate a weak coherence about which the opposing classical Bloch vectors precess [panels (i) and (ii)].
After application of the $\pi$-pulse to the 2nd collective spin this precession yields  net squeezing of the collective ensemble
about $\hat{y}$ [panel (iii)].

In contrast to OAT squeezing, the TSS dynamics are generated by a cavity field with an amplitude reduced by a factor of $\sqrt{N}$. This is a consequence of the field being induced by quantum fluctuations 
of the atomic coherence $\langle{\hat{a}\rangle}_{TSS} \sim \sqrt{\langle{\hat{S}_y^2}\rangle}\sim \sqrt{N}$. Regardless of this relatively weak cavity field, the TSS protocol can achieve a similar level of shearing relative to OAT. This is reconciled by understanding that in OAT the Bloch vector precesses at a rate 
$\propto N$ -- related to the cavity field amplitude $ \langle {\hat{a}\rangle}_{OAT} \sim \langle{\hat{S}^+ \rangle} \sim N$ -- about a rotation axis that is nearly aligned to said Bloch vector, 
up to fluctuations associated with atomic projection noise $\propto \sqrt{N}$. 
Conversely, in TSS the rotation rate is lower by a factor of $\sqrt{N}$ relative to OAT, however, the Bloch vectors associated with the individual spin ensembles are nearly perpendicular to the axis of rotation. 
Thus, in TSS the component of the Bloch vectors perpendicular to the axis of rotation is $\sqrt{N}$ larger than OAT, compensating for the $\sqrt{N}$ smaller cavity field compared to OAT.
The $\sqrt{N}$ larger component of the relevant Bloch vectors perpendicular to the axis of rotation compensates for the $\sqrt{N}$ smaller cavity field compared to OAT.

%

We now shift our focus to understand why the TSS squeezing is robust against collective emission. Whilst the reduced atomic coherence of the initial state leads to an associated
reduction in the rate of photon leakage from the cavity, $\propto \kappa N$, excess dissipative noise is still added to the $S_z$ quadrature of the final state at a rate identical to OAT, $\propto N^2 \Gamma t$. This is 
reconciled by noting that the $S_z$ quadrature of the measured state \emph{after} the rotation about $\hat{y}$ actually corresponds to the inversion difference $\hat{S}^z_1 - \hat{S}^z_2$ during the squeezing dynamics. 
In contrast to OAT, the dissipative noise in the measured $S_z$ quadrature is then not driven by the usual atomic coherence $\langle\hat{S}^+\hat{S}^-\rangle$ (i.e., cavity occupation $\langle \hat{a}^{\dagger}\hat{a}\rangle$),
but rather the \emph{differential} atomic coherence $\langle (\hat{S}^+_1 - \hat{S}^+_2)(\hat{S}^-_1 - \hat{S}^-_2)\rangle \sim N^2$. 
While this dissipative noise thus remains large, unlike OAT it now contributes predominantly to the \emph{anti-squeezed} quadrature of a \emph{phase-squeezed} state. 
We illustrate this contrast to OAT in Fig.~\ref{fig:SqFixGamma}(b). We emphasize that the reduced atomic 
coherence $\langle \hat{S}^+\hat{S}^- \rangle$ and suppression of superradiance remains an important ingredient for TSS: Superradiant decay of the atomic inversion would generate fluctuations 
in $\hat{S}_x$ and thus degrade correlations via the term $\propto (\hat{S}^x_1 - \hat{S}^x_2)^2$ in $\hat{\tilde{H}}$.

A perturbative treatment  leads to the following  approximate expression for the squeezing parameter (see SM):
\begin{equation}
 \xi^2_{\mathrm{TSS}} \approx \frac{1 + \Gamma N t}{2N\beta} + \frac{14}{9}\beta^2 . \label{eqn:TSSpert_superrad}
\end{equation}
The contrast to OAT is illustrated here by the suppression of the dissipative noise by the prefactor $\sim 1/(N\beta)$, which reflects that it is added predominantly to the anti-squeezed rather than
squeezed quadrature.
Importantly, this means that for TSS the optimal squeezing essentially remains limited only by the emergence of non-Gaussian corrections to the
distribution $\sim \beta^2$. Optimizing the squeezing respect to time  for fixed cavity parameters  Eq.~(\ref{eqn:TSSpert_superrad}) leads in this case to 
\begin{equation}
 \left. \xi^2_{\mathrm{TSS}} \right\vert_{\Gamma,\chi} \approx \frac{21^{1/3}}{2N^{2/3}} + \frac{7^{1/6}\Gamma}{3^{1/3}\chi N^{1/3}} . \label{eqn:TwoSpinSq_FixGam}
\end{equation} The key difference is that collective decoherence does not lead to a lower bound on squeezing as it does for conventional OAT. For large $N$ we thus find that TSS scales with atom number as
$\left. \xi^2_{TSS}\right\vert_{\Gamma,\chi} \propto N^{-1/3}$.

We directly compare the optimal squeezing generated by OAT and TSS protocols for the case  of $N=1000$ in Fig.~\ref{fig:SqFixGamma}. Results are based on numerical solution of the master equation taking
into account all relevant secondary effects, such as decay of the spin-length $|\langle \vec{S} \rangle|$ due to quantum fluctuations and non-collective terms $\propto (\hat{S}^x_1 - \hat{S}^x_2)^2$ in the
effective TSS Hamiltonian $\hat{\tilde{H}}$, so that the validity goes beyond the perturbative analysis of Eqs.~(\ref{eqn:OATpert_superrad}) and (\ref{eqn:TSSpert_superrad}). They confirm that, unlike OAT, suppression of
collective emission in TSS leads to squeezing limited by non-Gaussian corrections to the spin distribution at relatively long times. The strikingly different impact of superradiance in the schemes is illustrated in the inset,
where we plot the numerically obtained optimal squeezing as a function of $\Gamma/\chi$.

\noindent{\it Sensitivity to single-particle decoherence:} Instead of operating at fixed $\Gamma/\chi \equiv \kappa/\Delta_c$ one could in principle remove the detrimental effect of superradiance in OAT by operating at a
large cavity detuning $\Delta_c$. However, in reality under this condition the generation of squeezing will become sufficiently slow that other external and technical noise sources become
the limiting factors for metrological sensitivity. In this vein, we now include  relevant  single particle decoherence mechanisms which typically can be characterised in terms of the single particle
jump operators $\hat{L}^{s}_j = \sqrt{\gamma_{s}/2} \hat{\sigma}^-_j$ (describing, e.g., spontaneous emission or Raman light scattering) and $\hat{L}^{el}_j = \sqrt{\gamma_{el}/8}\hat{\sigma}^z_j$ (describing, e.g.,
Rayleigh scattering and dephasing from stray fields or collisions).

By treating single-particle and collective emission perturbatively (see SM) we obtain the approximate expressions
\begin{eqnarray}
 \xi^2_{\mathrm{TSS},\gamma_s} \approx \frac{1+\Gamma N t}{2N\beta} + \gamma_{s} t , \\
 \xi^2_{\mathrm{OAT},\gamma_s} \approx \frac{1}{2N\beta} + \Gamma N t + \gamma_s t ,
\end{eqnarray}
where we ignore the terms $\propto \beta^2$ as irrelevant compared to the dissipative contribution.
The clear difference here is that the squeezing achievable via OAT is limited by \emph{both} collective and single-particle emission, whereas TSS suppresses the collective component. 
One then expects that squeezing can be generated on faster time-scales with TSS to minimize single-particle decoherence, as collective dissipative noise is not the most relevant  limitation.
This is supported by optimising the achievable squeezing by varying the cavity detuning $\Delta_c$, yielding the superior
$\xi^2_{\mathrm{TSS}}\vert_{\gamma_s} \approx \sqrt{24/(N\eta_s)}$ when compared to $\xi^2_{\mathrm{OAT}}\vert_{\gamma_s} \approx 6(N\eta_s)^{-1/3}$ where $\eta_s = 4g^2/(\kappa\gamma_s)$ is an effective 
cavity co-operativity. The optimal detuning for TSS
is closer to the cavity resonance by a factor of $\sim (\eta_s N)^{1/4}$ and occurs on a shorter overall time-scale than OAT, confirming expectations (see SM for details).
We note that the scaling achievable with TSS is equivalent to that
predicted by introducing an additional drive term in Ref.~\cite{Vuletic2017}. However, the improvement in that case is only a byproduct of enhancing the rate at which the squeezing dynamics occur, at the cost
of adding further systematic effects generated by the additional drive.
Whether a driving protocol can be similarly implemented to accelerate TSS will be addressed in a follow-up work
as well as applications to more general interferometry schemes \cite{SchleierSmith_HeisenbergLimit_2016,Kasevich2016,Nolan2017}.

However, the robustness of TSS to collective emission comes at a tradeoff to increased sensitivity to single-particle dephasing. This decoherence intuitively adds
excess noise to the squeezed quadrature of the phase-squeezed state, whereas for OAT it only contributes to the anti-squeezed quadrature. We find the time evolution of the squeezing for $\gamma_{el}t \ll 1$
is approximately (see SM)
\begin{eqnarray}
 \xi^2_{\mathrm{TSS},\gamma_{el}} \approx \frac{1+\Gamma N t}{2N\beta} + \gamma_{el} t , \\
 \xi^2_{\mathrm{OAT},\gamma_{el}} \approx \frac{1 + 2\gamma_{el}t}{2N\beta} + \Gamma N t .
\end{eqnarray}
Here, the suppression of collective emission in TSS is of reduced benefit over OAT, as the role of single-particle dephasing and collective emission is interchangeable between the two schemes. This is reflected
by optimising with respect to the cavity detuning, which yields an identical result $\xi^2 \sim (N\eta_{el})^{-1/2}$ for both protocols (see SM) where $\eta_{el} = 4g^2/(\kappa\gamma_{el})$. Consequently, 
the superiority of TSS over OAT, driven by the ability to suppress collective dissipative noise, is most prominent in the limit where single-particle decoherence is dominated by spontaneous emission, $\gamma_s \gg \gamma_{el}$.
Such a regime is relevant as $\gamma_s$ and $\Gamma$ are fundamental sources of decoherence, due to the finite transition linewidth and engineering of the squeezing Hamiltonian respectively,
whereas $\gamma_{el}$ is a technical barrier.


\noindent{\it Conclusion:}  Whilst the generation of quantum correlations and entanglement which are protected against decoherence is of broad interest to quantum enhanced technologies, to date, measurement improvements from 
deterministic generation of many-body states have been small and proof-of-principle in nature. The TSS protocol proposed here is  intrinsically robust
against the detrimental effects of superradiance and thus  particularly useful for the next-generation of quantum enhanced optical atomic clocks. Its implementation could open a path to
deliver significant gains to sensors with real-world applications.


\acknowledgments
The authors acknowledge fruitful discussions with J.~Young, J.~Viennot and A.~Gorshkov.
A.~M.~R acknowledges support from Defense Advanced Research Projects Agency (DARPA) and Army Research Office grant W911NF-16-1-0576, NSF grant PHY1521080, JILA-NSF grant PFC-173400, and the
Air Force Office of Scientific Research and its Multidisciplinary University Research Initiative grant FA9550-13-1-0086. Financial support from NIST is also acknowledged.

\bibliography{TwoSpinSqueezing_refs}

\newpage 

\onecolumngrid
\vspace{\columnsep}
\begin{center}
\textbf{\large Supplemental Material: Robust spin squeezing via photon-mediated interactions on an optical clock transition}
\end{center}
\vspace{\columnsep}
\twocolumngrid

\setcounter{equation}{0}
\setcounter{figure}{0}
\setcounter{table}{0}
\setcounter{page}{1}
\makeatletter
\renewcommand{\theequation}{S\arabic{equation}}
\renewcommand{\thefigure}{S\arabic{figure}}
\renewcommand{\bibnumfmt}[1]{[S#1]}
\renewcommand{\citenumfont}[1]{S#1}

\section{Atom-light interaction and effective spin model}
In the main text we assume the dynamics of the CQED system can be reduced to an effective spin model describing the atomic degrees of freedom, with the cavity field slaved to the atomic coherence. 
Here, we present a more detailed derivation that establishes this approximation. 

Recapping the main text for completeness, we consider a system of $N$ atoms trapped in a standing-wave optical lattice which is supported by an optical cavity. The cavity field 
couples the ground and excited clock states of an atom with single-photon Rabi frequency $2g_{\alpha}$, where the subscript $\alpha$ denotes the dependence on the internal hyperfine ($m_F$) degree of freedom. 
We describe the atomic degree of freedom using collective spin operators 
$\hat{S}^{x,y,z}_{j,\alpha} \equiv \sum_{n\in j} \hat{\sigma}^{x,y,z}_{n,\alpha}/2$ where $\hat{\sigma}^{x,y,z}_{j,\alpha}$ denote the conventional Pauli matrices. Here, 
the subscript $j$ of the collective spin refers to the spatial lattice site, and the summation over $n$ refers to atoms with identical lattice position. 

Most generally, the dynamics of the coupled  atom-light system is described by a master equation for the density matrix, $\hat \rho$,
\begin{equation}
 \frac{d\hat{\rho}}{dt} = -\frac{i}{\hbar}\left[\hat{H}_{\mathrm{AL}},\hat{\rho}\right] + \mathcal{L}_c[\hat{\rho}],  \label{eqn:AL_master_eqn}
\end{equation}
Here, the Hamiltonian describing the atom-light coupling is
\begin{equation}
 \hat{H}_{\mathrm{AL}} = \hbar\Delta_c\hat{a}^{\dagger}\hat{a} + \hbar \sum_{j,\alpha} g_{j,\alpha} \left( \hat{a}^{\dagger}\hat{S}^-_{j,\alpha} + \hat{a}\hat{S}^+_{j,\alpha} \right) , \label{eqn:Ham_AL}
\end{equation}
where $\Delta_c$ characterizes the relative detuning of the cavity field from the atomic transition and we have introduced an additional spatial dependence $j$ for the atom-light coupling. In principle there 
may be additional single-particle terms $\propto \hat{\sigma}^z_i$ describing inhomogeneities of the atomic transition frequencies. However, we ignore this in the following and mention it only to point 
out that any detrimental effects of such broadening-type terms may in principle be removed in the experiment by spin-echo. The Lindblad term 
\begin{equation}
 \mathcal{L}_c[\hat{\rho}] = \frac{\kappa}{2}\left( 2\hat{a}\hat{\rho}\hat{a}^{\dagger} - \hat{a}^{\dagger}\hat{a}\hat{\rho} - \hat{\rho}\hat{a}^{\dagger}\hat{a} \right) .
\end{equation}
describes photon loss from the cavity with power decay rate $\kappa$.

Operating in the bad cavity limit $\kappa \gg \gamma$ and assuming that cavity loss occurs at a much faster rate than the atomic dynamics, $\kappa \gg g$, we may adiabatically eliminate the cavity mode. 
This leads to a slaving of the cavity field to the atomic coherence,
\begin{equation}
 \hat{a}(t) \equiv \frac{2}{2\Delta_c + i\kappa} \sum_{j,\alpha} g_{j,\alpha} \hat{S}^{-}_{j,\alpha} . \label{eqn:cavity_slaving}
\end{equation}
In turn this simplifies the model of Eq.~(\ref{eqn:AL_master_eqn}) to a master equation for the reduced density matrix $\hat{\rho}_s$ of the spins,
\begin{equation}
 \frac{d\hat{\rho}_s}{dt} = -\frac{i}{\hbar} \left[ \hat{H}_{\mathrm{eff}}, \hat{\rho}_s \right] + \hat{L}[\hat{\rho}_s] , \label{eqn:spin_master_eqn}
\end{equation}
Here, the effective Hamiltonian is
\begin{equation}
 \hat{H}_{\mathrm{eff}} = \hbar \sum_{j,l,\alpha,\beta} \chi_{j,\alpha,l,\beta} \hat{S}^+_{j,\alpha} \hat{S}^-_{l,\beta} ,
\end{equation}
where $\chi_{j,\alpha,l,\beta} \equiv 4 g_{j,\alpha} g_{l,\beta} \Delta_c/(4\Delta_c^2 + \kappa^2)$ is the strength of the elastic interaction. This is accompanied by a 
dissipative contribution which describes collective emission into the cavity mode, 
\begin{eqnarray}
 \hat{L}[\hat{\rho}_s] = \sum_{j,l,\alpha,\beta} \frac{\sqrt{\Gamma_{j,\alpha}\Gamma_{l,\beta}}}{2} \Big( 2\hat{S}^-_{j,\alpha}\hat{\rho}_s\hat{S}^+_{l,\beta} \notag \\
  - \hat{S}^+_{j,\alpha}\hat{S}^-_{l,\beta}\hat{\rho}_s - \hat{\rho}_s\hat{S}^+_{j,\alpha}\hat{S}^-_{l,\beta} \Big)
\end{eqnarray}
where $\Gamma_{j,\alpha} = 4g^2_{j,\alpha}\kappa/(4\Delta_c^2 + \kappa)$.

Throughout the main text we assume the atom-light coupling is spatially uniform, $g_{j,\alpha} \equiv g_{\alpha}$, which requires that the wavelength of the standing wave lattice and cavity field are 
commensurate. This assumption allows the simplification $\chi_{j,\alpha,l,\beta} = \chi_{\alpha,\beta}$ and $\Gamma_{j,\alpha} = \Gamma_{\alpha}$. In the case of the OAT scheme, where only 
one internal degree of freedom ($m_F$ state) is involved, the Hamiltonian is then simplified to $\hat{H}_{\mathrm{eff}} \equiv \chi \hat{S}^+\hat{S}^-$ and the Lindblad jump operator 
becomes $\hat{L}_{\Gamma} \equiv \sqrt{\Gamma/2}\hat{S}^-$, for $\chi \equiv \chi_{\alpha,\alpha}$ and $\Gamma \equiv \Gamma_{\alpha}$. 
On the other hand, as the TSS scheme involves a pair of internal states it requires an additional assumption. To achieve a uniform coupling, with respect to the internal degrees of 
freedom, we require that the atomic ensembles realizing the collective spins occupy a symmetric pair of hyperfine states such that $m_F = \pm m$. 

%
%

\section{Optimal spin squeezing in the OAT and TSS schemes \label{sec:IdealSqueezing}}
In this section we outline some analytic results describing the achievable spin-squeezing in the OAT and TSS schemes in the presence of both collective and single-particle dissipation.

\subsection{Exact solution of one-axis twisting \label{sec:ExactOAT}}
The dynamics of a single coherent spin state under $\hat{H} = \chi \hat{S}^+\hat{S}^-$ can be solved analytically, by noting that we can rewrite the Hamiltonian as $\hat{H} \equiv \hat{S}^2 - \hat{S}_z^2 + \hat{S}_z$. 
Ignoring the single-particle rotation, and using that $\hat{S}^2$ commutes with the one-axis twisting term, we have the simpler Hamiltonian one-axis twisting Hamiltonian $\hat{H}_{\mathrm{OAT}} = \chi \hat{S}^2_z$. 
Dynamics under this Hamiltonian is exactly solvable, and the particular case of spin squeezing is covered extensively in the literature. We refer the interested reader to 
Refs.~\cite{Ueda_SpinSqueezing_1993,FossFeig2013,FossFeig2013NJP} in particular. In this section, we simply present the relevant results for the case of ideal OAT in the absence of any decoherence. 

We begin with the minimal (maximal) quadrature variance 
$V_- \equiv \mathrm{min}_{\psi}[ \langle (\delta \hat{S}_{\psi})^2 \rangle ]$ ($V_+ \equiv \mathrm{max}_{\psi}[ \langle (\delta \hat{S}_{\psi})^2 \rangle ]$), which is given by 
\begin{equation}
 V_{\pm} = \frac{1}{2}\left[ \mathcal{C}(t) \pm \sqrt{\mathcal{A}(t)^2 + \mathcal{B}(t)^2} \right] , \label{eqn:UedaSqQuad}
\end{equation}
where 
\begin{eqnarray}
 \mathcal{A}(t) & \equiv & \langle \left(\delta \hat{S}_y\right)^2 - \left(\delta \hat{S}_z\right)^2 \rangle , \label{eqn:A_OAT} \\
 \mathcal{B}(t) & \equiv & \langle \hat{S}_y\hat{S}_z + \hat{S}_z \hat{S}_y \rangle , \label{eqn:B_OAT} \\
 \mathcal{C}(t) & \equiv & \langle \left(\delta \hat{S}_y\right)^2\rangle + \langle \left(\delta \hat{S}_z\right)^2 \rangle . \label{eqn:C_OAT} 
\end{eqnarray}
Here, the relevant correlations are given by \cite{Ueda_SpinSqueezing_1993},
\begin{widetext}
\begin{eqnarray}
 \langle \left(\delta \hat{S}_y\right)^2 \rangle = \frac{S}{2} + \frac{S}{2}\left( S - \frac{1}{2}\right)\left[1 - \mathrm{cos}^{2S-2}(2\tau)\right] , \\
 \langle \left(\delta \hat{S}_z\right)^2 \rangle = \frac{S}{2} , \quad
 \langle \hat{S}_y\hat{S}_z + \hat{S}_z \hat{S}_y \rangle = 2S\left(S - \frac{1}{2}\right) \mathrm{sin}(\tau)\mathrm{cos}^{2S-2}(\tau) .
\end{eqnarray}
\end{widetext}
for $\tau = \chi t$ and $S = N/2$. Defining $\nu = (1/2)\mathrm{arctan}(\mathcal{B}/\mathcal{A})$, the maximum variance is for $\psi = \nu$ and the minimum variance when $\psi = \pi/2 - \nu$. 
For $S \gg 1$ decay of the effective spin length can be ignored, e.g, $|\langle \vec{S} \rangle| \equiv |\langle \hat{S}_x \rangle| \approx S$, the time evolution of the 
squeezing parameter is given by $\xi^2 \approx V_-/(S/2)$. 

A more insightful expression for the squeezing can be obtained by expanding Eqs.~(\ref{eqn:A_OAT})-(\ref{eqn:C_OAT}) perturbatively in the small parameter $\beta = S\tau^2 \ll 1$. Although this analysis 
has been presented previously in the literature, (see, e.g., Ref.~\cite{Ji2013}), for clarity of our later calculations dealing with TSS and decoherence, we outline the essential approach here. 

First, to $\mathcal{O}(\beta^3)$ we obtain
\begin{eqnarray}
 \mathcal{A} \approx \frac{S^2}{2}\left( 4\beta - 8\beta^2 + \frac{32}{3}\beta^3 \right) , \\
 \mathcal{C} \approx S + \frac{S^2}{2}\left( 4\beta - 8\beta^2 + \frac{32}{3}\beta^3 \right) , 
\end{eqnarray}
whilst $\mathcal{B} \approx 2S^2\tau e^{-\beta}$ and thus
\begin{equation}
 \mathcal{B}^2 \approx 4S^3\beta \left( 1 - 2\beta + 2\beta^2 \right) .
\end{equation}
Here, we have used that $\mathrm{cos}(\tau) \approx e^{-\tau^2/2}$ for $\tau \ll 1$. Also, for each expression we have taken the leading order contribution in $S$ for each coefficient of $\beta^n$. 

We proceed by calculating the product of the squeezing and anti-squeezed quadratures,
\begin{eqnarray}
 V_{+}V_{-} & \equiv & \frac{1}{4}\left[ \left( \mathcal{C} + \mathcal{A} \right)\left( \mathcal{C} - \mathcal{A} \right) - \mathcal{B}^2 \right] , \\
 & \approx & \frac{S^2}{4}\left[ 1 + \frac{8}{3}S\beta^3 \right] . 
\end{eqnarray}
Next, we obtain an approximation expression for $V_+$ to leading order in $\beta$ by assuming that for $S^2\tau > 1$ we have $\mathcal{A} > \mathcal{B}$ and thus we 
approximate $\sqrt{\mathcal{A}^2 + \mathcal{B}^2} \approx \mathcal{A}$. This leads to 
\begin{eqnarray}
 V_{+} & = & \frac{1}{4}\left[ \mathcal{C} + \mathcal{A} \right] , \\
 & \approx & 2S^2 \beta .
\end{eqnarray}
Here we have assumed $2S^2\beta \gg S$ (valid as we assume this quadrature is highly anti-squeezed with respect to the initial state).

It is then straightforward to obtain the squeezing parameter by the relation
\begin{eqnarray}
 \xi^2_{OAT} = \frac{V_+V_-}{V_+}\frac{2}{S} \approx \frac{1}{2N\beta} + \frac{2}{3}\beta^2 . \label{eqn:OATpert}
\end{eqnarray}
The first term of this expression describes the squeezing of the state, whilst the second term describes a correction due to the state becoming non-Gaussian (i.e. `oversqueezing'). 
In the case of OAT this correction occurs as the squeezed distribution begins to probe the curvature of the Bloch sphere. 

The optimal squeezing as a function of system size $N$ can be found by minimizing Eq.~(\ref{eqn:OATpert}) with respect to the evolution time, yielding
\begin{equation}
 \xi^2_r \approx (9/8)^{1/3}N^{-2/3} \label{eqn:OAT_IdealScaling}
\end{equation}
for $\tau_{\mathrm{opt}} = 3^{1/6}N^{-2/3}$.

\subsection{Semi-classical treatment of two-spin squeezing \label{sec:TSS_Ideal}}
In this section we present an analytic treatment of two-spin squeezing using a semi-classical approximation of the dynamics. In particular, we solve the quantum dynamics using the truncated Wigner approximation, which 
allows us to take into account the non-collective effects of the TSS Hamiltonian. The resulting expression allows us to rigorously quantify and understand the effects of the non-collective terms, and we find they have limited 
consequences for the squeezing dynamics, justifying our discussion in the main text. 

For simplicity, our solution considers the dynamics of the system in the frame of the initial back-to-back state prepared along $\pm\hat{x}$, $\vert \psi_0 \rangle \equiv \vert N/4 \rangle_x \otimes \vert -N/4 \rangle_x$, 
with the Hamiltonian given by $\hat{H} = \chi \hat{S}^+\hat{S}^-$. Here, we remind the reader that the operators should be interpreted as $\hat{S}_{\alpha} \equiv \hat{S}^{\alpha}_1 + \hat{S}^{\alpha}_2$ where 
the subscript denotes the individual collective spins. 

Solving the dynamics according to the truncated Wigner approximation consists of two steps. First, introducing the c-number variables $S_{\alpha} \equiv \langle \hat{S}_{\alpha}\rangle$, we analytically solve the 
mean-field equations of motion for the system. Second, we obtain the appropriate symmetrically-ordered quantum expectation values by averaging over initial conditions with respect to the Wigner quasi-probability 
distribution $W(S_x,S_y,S_z)$ \cite{Polkovnikov_PhaseSpace_2010}. For example,
\begin{multline}
 \langle \hat{S}_y(t)\hat{S}_z(t) + \hat{S}_z(t)\hat{S_y}(t) \rangle \\  \equiv 2\int dS_x dS_y dS_z~ W(S_x,S_y,S_z) S_y(t) S_z(t) . 
\end{multline}
For the back-to-back state under consideration the Wigner function factorizes in terms of the constituent collective spins, 
\begin{equation}
 W(S_x, S_y, S_z) \equiv W_+(S^x_1,S^y_1,S^z_1) W_-(S^x_2,S^y_2,S^z_2)
\end{equation}
where 
\begin{equation}
 W_\pm(S^x_i,S^y_i,S^z_i) = \frac{2}{\pi N} \delta\left(S^x_i \pm \frac{N}{4}\right) e^{-\frac{2}{N}\left[ (S^y_i)^2 + (S^z_i)^2 \right]} . 
\end{equation}

We begin with the relevant equations of motion for the c-number variables:
\begin{eqnarray}
 \frac{dS^+_1}{dt} = -i2\chi\left(S^+_1 + S^+_2\right)S^z_1 , \\
 \frac{dS^+_2}{dt} = -i2\chi\left(S^+_1 + S^+_2\right)S^z_2 , \\
 \frac{dS^z_1}{dt} = -i\chi\left(S^+_1S^-_2 - S^-_1S^+_2\right) , \\
 \frac{dS^z_2}{dt} = i\chi\left(S^+_1S^-_2 - S^-_1S^+_2\right) .
\end{eqnarray}
The equations are most easily solved by the introduction of sum and difference variables, $S_{\alpha} \equiv S^{\alpha}_1 + S^{\alpha}_2$ and $\Delta_{\alpha} \equiv S^{\alpha}_1 - S^{\alpha}_2$. 
As the total magnetization is conserved, $S_z(t) \equiv S_z(0)$, a solution for the atomic coherence is straightforward and we obtain $S^+(t) = S^+(0) e^{-2i S_z(0) \tau}$. The solution for the 
magnetization difference $\Delta_z(t)$ is more sophisticated, but essentially reduces to the solution of a second-order differential equation of the form $\ddot{\Delta_z} = a + b\Delta_z$, where $a$ and $b$ are functions of 
the initial conditions. Exact solution of the differential equation is straightforward, but we do not present the resulting expression for $\Delta_z(t)$ here as it is lengthy and not insightful. 

With the analytic expressions of $S^+(t)$ and $\Delta_z(t)$ in hand, we proceed by obtaining perturbative expressions for the quantities 
 \begin{eqnarray}
 \mathcal{A}(t) \equiv \langle \left(\delta \hat{S}_y\right)^2 - \left(\delta \hat{\Delta}_z\right)^2 \rangle \label{eqn:A_TSS} \\
 \mathcal{B}(t) \equiv \langle \hat{S}_y\hat{\Delta}_z + \hat{\Delta}_z \hat{S}_y \rangle  \label{eqn:B_TSS} \\
 \mathcal{C}(t) \equiv \langle \left(\delta \hat{S}_y\right)^2\rangle + \langle \left(\delta \hat{\Delta}_z\right)^2 \rangle \label{eqn:C_TSS} .
\end{eqnarray}
as per the OAT treatment of the prior subsection. Here, we have substituted $\hat{S}_z \to \hat{\Delta}_z$ as the relevant quantity in the frame of the back-to-back spins, 
i.e. before squeezing is measured in the final frame where one of the ensembles is rotated by a $\pi$-pulse about $\hat{y}$.

After expansion of Eqs.~(\ref{eqn:A_TSS})-(\ref{eqn:C_TSS}) to $\mathcal{O}(\beta^3)$ in the small parameter $\beta = S\tau^2 \ll 1$ and following the procedure of the previous section identically,
we obtain a perturbative expression for the squeezing parameter,
\begin{eqnarray}
 \xi^2_{TSS} & \approx & \frac{1}{2N\beta} + \frac{14}{9}\beta^2 . \label{eqn:TSSpert}
\end{eqnarray}
Here, the key result is that the term $\mathcal{O}(\beta^2)$, which describes the onset of non-gaussian corrections to the distribution, has an enhanced prefactor relative to OAT. 
This indicates that the non-collective terms become relevant when the squeezed distribution begins to probe the curvature of the Bloch sphere, and thus leads to the minimum squeezing occuring slightly earlier.
Such an effect is consistent with the simplistic argument of the main text, which emphasizes that the contribution of the non-collective terms should be small at short times (i.e., the time-scale of the squeezing dynamics), 
as the initial state is an eigenstate of $\hat{S}_x$ with $\langle \hat{S}_x \rangle = 0$. 

Quantitatively, we find these corrections lead to a minor decrease in the optimal squeezing time $\tau_{\mathrm{opt}} = (3^{2}/7)^{1/6}N^{-2/3}$, and the optimal squeezing is
\begin{equation}
 \xi^2_{TSS,\mathrm{opt}} \approx \frac{21^{1/3}}{2N^{2/3}} . \label{eqn:TSS_IdealScaling}
\end{equation}
Thus, squeezing is merely reduced by approximately $1.2$~dB relative to OAT. 

\begin{figure}
 \includegraphics[width=8cm]{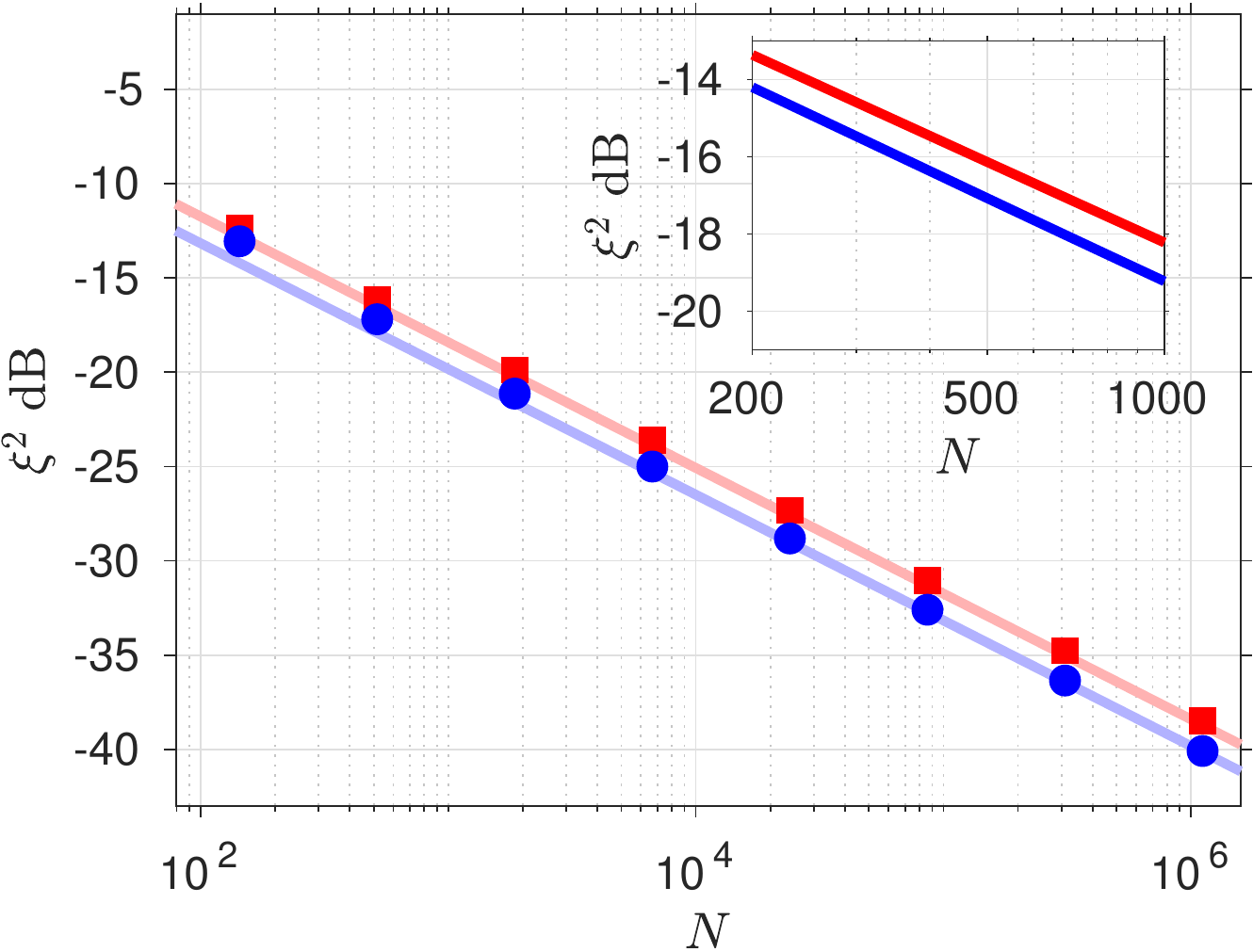}
 \caption{Scaling of ideal OAT and TSS squeezing with system size $N$. (Main) Comparison of analytic expressions for OAT [Eq.~(\ref{eqn:OAT_IdealScaling}), faded blue line] and TSS [Eq.~(\ref{eqn:TSS_IdealScaling}), faded red line], with 
 those obtained by full numerical solution using TWA (OAT - blue circles, TSS - red squares). (Inset) Comparison of OAT (blue) and TSS (red) from exact diagonalization (see text for discussion of truncation in TSS solution). Squeezing 
 is different only by an approximately unchanged shift of $\sim 1$~dB across the range of $N$ calculated, consistent with analytic predictions of Eqs.~(\ref{eqn:OAT_IdealScaling}) and (\ref{eqn:TSS_IdealScaling}).}
 \label{fig:OAT_TSS_Scaling}
\end{figure}

Our analytic results are confirmed by full numerical calculations, presented in Fig.~S\ref{fig:OAT_TSS_Scaling}. In particular, the perturbative expression is validated 
by TWA calculations for large $N$, whilst the prediction of TSS following OAT scaling up to a small prefactor correction is consistent with a solution of the dynamics using exact diagonalization for 
small system sizes $N \leq 1000$ and truncating the relevant Hilbert space. This truncated exact solution is based on solving the problem in the rotated frame, 
$\hat{\tilde{H}} = \chi(\hat{S}^x_1 - \hat{S}^x_2)^2 + \chi(\hat{S}^y_1 + \hat{S}^y_2)^2$. 
Here, the initial state is a collective spin in the fully symmetric manifold $S=N/2$. However, the non-collective term of $\hat{\tilde{H}}$ can lead to population of other Dicke manifolds $S=N/2-1,...0$. 
Nevertheless, as the dynamics in the rotated frame is predominatly driven by the fully collective term $\hat{S}_y^2$ at short times, then in principle only a few manifolds $S=N/2,N/2-1,...N/2-n$ for $n \ll N/2$ will 
be relevant, which allows us to truncate the Hilbert space. Indeed generically $n = 4$ is sufficient for the systems considered. 
This truncation reduces the numerical complexity of the problem from simulating a Hilbert space of $(N/2 + 1)^2$ to $\sim (N+1)n/2$

The calculations presented in Fig.~S\ref{fig:OAT_TSS_Scaling} fully justify our physical discussion of TSS in the main text, based on the simplified Hamiltonian $\hat{H} \approx \chi\hat{S}_y^2$.

\section{Effects of collective emission \label{sec:CollDiss}}
In this section, we expand on the idealized results of the previous section and include the effects of collective emission, described by the Lindblad jump operator $\hat{L}_{\Gamma} \equiv \sqrt{\Gamma/2}\hat{S}^-$, 
which is intrinsically present in the cavity-QED realization of both OAT and TSS schemes using the optical clock transition. 

\subsection{One-axis twisting with collective emission}
Our treatment of collective emission for the OAT scheme follows that presented in Ref.~\cite{Vuletic2017}. For clarity, we summarize the approach and relevant approximations here. 

Collective emission does not distinguish between atoms, and preserves the total spin quantum number S. For the case of a coherent spin state with maximum initial coherence, the leakage of a photon from the 
cavity induces a variance in the atomic spin distribution of the form $\langle (\delta\hat{S}_z)^2 \rangle = (N/2)\mathrm{tanh}(N\Gamma t/2)[1 - \mathrm{tanh}(N\Gamma t/2)]$ \cite{Gross1982}. 
For weak decoherence, $N\Gamma t \ll 1$, we assume that the dissipative dynamics can be treated 
independently of the one-axis twisting dynamics and that additional fluctuations generated by collective emission can be added in quadrature to the squeezed state.

Under this approximation, and further noting that the squeezed quadrature $V_-$ is approximately aligned along $S_z$ ($\psi \rightarrow \pi/2$ for $N \gg 1$), the perturbative expression for the squeezing parameter 
is then modified,
\begin{equation}
 \xi^2_{\mathrm{OAT},\Gamma} \approx \frac{1}{2N\beta} + \frac{2}{3}\beta^2 + N\Gamma t . \label{eqn:OATGamma_pert}
\end{equation}
For $\Gamma \neq 0$ the squeezing rapidly becomes limited by the dissipative contribution. The optimal squeezing time can then be obtained by minimizing $1/(2N\beta) + N\Gamma t$, 
yielding $t_{\mathrm{opt}} \approx [2/(N^3\chi^2\Gamma)]^{1/3}$ and the best achievable squeezing 
\begin{equation}
 \left. \xi^2_{\mathrm{OAT}} \right\vert_{\Gamma,\chi} \approx \frac{2}{3^{2/3}} \left( \frac{\Gamma}{\chi} \right)^{2/3} , \label{eqn:OATGamma_bound}
\end{equation}
which is independent of atom number $N$.

\subsection{Two-spin squeezing with collective emission \label{sec:TSS_CollEmit}}

Establishing an analytic model to describe TSS in the presence of collective emission is a more difficult task. For simplicity, in this section we consider the dynamics of the system in the rotated 
frame, where the Hamiltonian and Lindblad term become
\begin{eqnarray}
 \hat{\tilde{H}} \approx \chi(\hat{S}^x_1 - \hat{S}^x_2)^2 + \chi(\hat{S}^y_1 + \hat{S}^y_2)^2 , \\
 \hat{\tilde{L}} \equiv \sqrt{\Gamma/2}(\hat{S}^x_1 - \hat{S}^x_2) - i\sqrt{\Gamma/2}(\hat{S}^y_1 + \hat{S}^y_2) ,
\end{eqnarray}
respectively, and the initial state is $\vert \psi_0 \rangle \equiv \vert N/4 \rangle_x \otimes \vert N/4 \rangle_x$.

\begin{figure}
 \includegraphics[width=8cm]{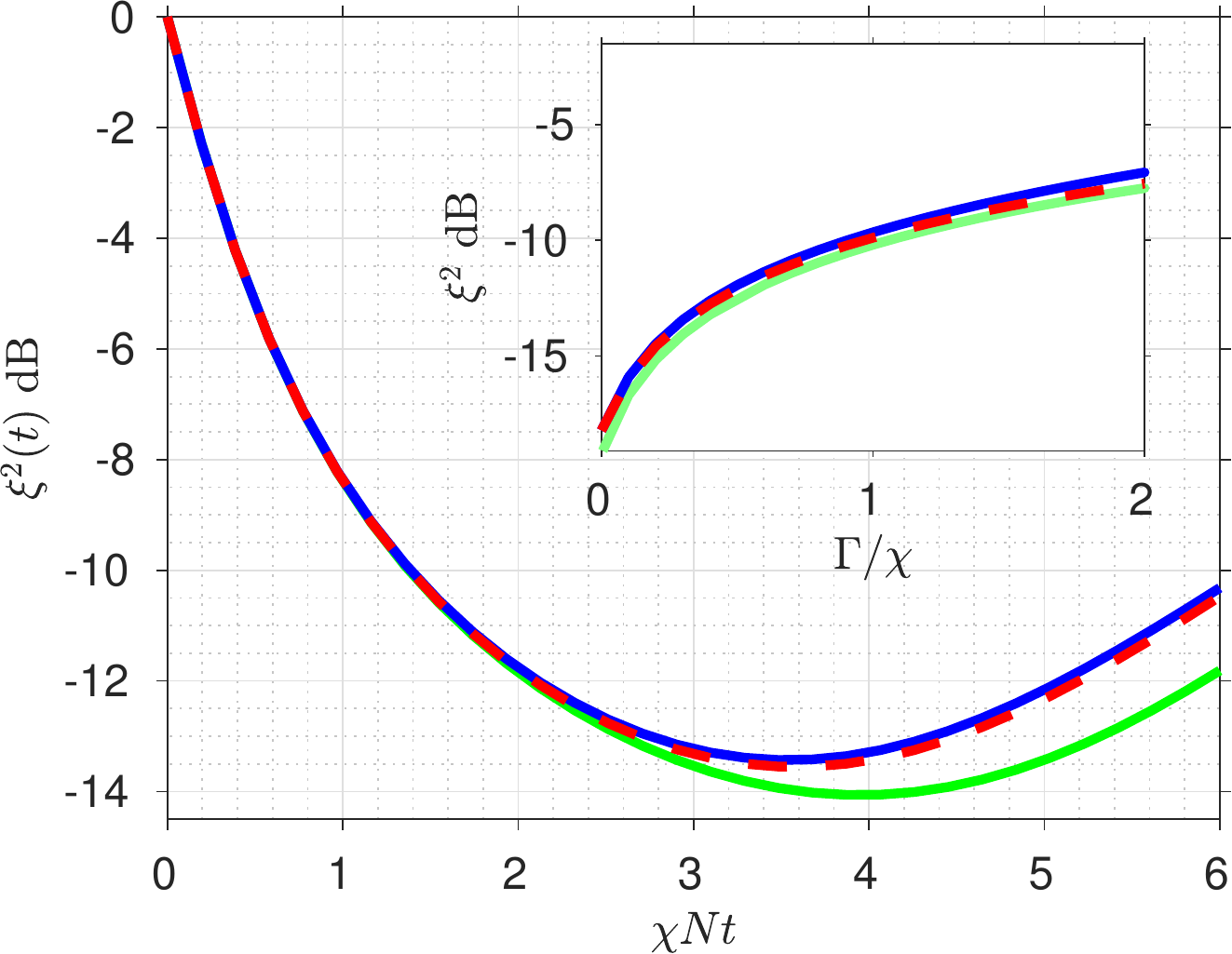}
 \caption{Comparison of squeezing calculated with: complete $\hat{\tilde{H}}$ and $\hat{\tilde{L}}$ (blue data, see text for details), complete $\hat{\tilde{H}}$ and approximate 
 $\hat{\tilde{L}} \approx \sqrt{\Gamma/2}\hat{S}_y$ (dashed red data), and approximate $\hat{\tilde{H}} \approx \chi S_y^2$ and $\hat{\tilde{L}} \approx \sqrt{\Gamma/2}\hat{S}_y$ (green data). (Inset) Optimal squeezing 
 as a function of dissipation strength $\Gamma/\chi$ (lines as previous). For $\Gamma/\chi \lesssim 1$ the predominant correction to the squeezing dynamics is from the non-collective term in the Hamiltonian $\hat{\tilde{H}}$, 
 whilst $\hat{\tilde{L}} \approx \sqrt{\Gamma/2}\hat{S}_y$ is an excellent approximation of the dissipation. All calculations are for $N = 1000$.}
 \label{fig:CheckDissTSS}
\end{figure}

The result Eq.~(5) of the main text is obtained by making a pair of approximations with respect to the Hamiltonian and Lindblad terms. First, we assume that the squeezing generated by TSS is well captured by the 
simplified Hamiltonian $\hat{\tilde{H}} \approx \chi\hat{S}_y^2$. This approximation is well supported by the results of Sec.~\ref{sec:IdealSqueezing}. Secondly, we assume that the Linblad term can be similarly 
simplified to $\hat{\tilde{L}} \approx \sqrt{\Gamma/2}\hat{S}_y$. This second assumption is justified by 
numerical simulations of systems up to $N\sim 1000$, examples of which are shown in Fig.~S\ref{fig:CheckDissTSS}, together with qualitative physical arguments discussed below. 

Under these approximations, the problem reduces to that of one-axis twisting with collective dephasing. This model is tractable as the dephasing commutes with the Hamiltonian dynamics and has been solved analytically 
in Ref.~\cite{Ji2013}. A perturbative expression for the squeezing parameter is obtained identically to Sec.~\ref{sec:IdealSqueezing}, 
\begin{equation}
 \xi^2 \approx \frac{1 + \Gamma N t}{2N\beta^{\prime}} + \frac{2}{3}\beta^{\prime 2} , \label{eqn:TSSopt_dephase}
\end{equation}
where $\beta^{\prime} \equiv \beta + \Gamma t/2$. In the main text we assume that $(\Gamma/\chi)\tau \ll S\tau^2$, such that $\beta^{\prime} \approx \beta$ and 
\begin{eqnarray}
 \xi^2_{\mathrm{TSS},\Gamma} & \approx & \frac{1 + \Gamma N t}{2N\beta}  + \frac{14}{9}\beta^2 . \label{eqn:TSSpert_superrad}
\end{eqnarray}
Here, and in the main text, we artificialy alter the prefactor of the term $\propto \beta^2$, from $2/3$ to $14/9$, 
as a crude correction to better include the effect of the non-collective terms of the Hamiltonian (although this does not 
change any physics qualitatively). As discussed in the main text, the fact that squeezing becomes predominantly limited only by non-Gaussian corrections means that the best achievable squeezing is given by
\begin{equation}
 \left. \xi^2_{\mathrm{TSS}} \right\vert_{\Gamma,\chi} \approx \frac{21^{1/3}}{2N^{2/3}} + \frac{7^{1/6}\Gamma}{3^{1/3}\chi N^{1/3}} , \label{eqn:TSSGamma_bound}
\end{equation}
which asymptotically scales as $N^{-1/3}$ for large $N$.

Equation (\ref{eqn:TSSpert_superrad}) is also consistent with a far more crude treatment of the decoherence, which more clearly illuminates the difference between the OAT and TSS protocols. 
Following a similar procedure to the OAT case, we can consider the effects of the collective decoherence as an independent and perturbative correction to the squeezing expression. In particular, 
for $\Gamma t \ll N^{-1/2}$ the dominant dissipative contribution from $\hat{L} = \sqrt{\Gamma/2}(\hat{S}^x_1 - \hat{S}^x_2) - i\sqrt{\Gamma/2}(\hat{S}^y_1 + \hat{S}^y_2)$ is additional noise along $\hat{S}_z$, 
$\langle (\delta \hat{S}_z )^2 \rangle \approx N^2\Gamma t$. However, as discussed in the main text, as TSS leads to a phase-squeezed state with $V_-$ aligned closely to $\hat{y}$ this implies that, unlike OAT, 
the excess noise from collective dissipation contributes predominantly to the anti-squeezed quadrature. However, as the squeezing angle is not quite zero, the squeezed quadrature still admits some small component 
of the dissipative noise. In particular, adding the dissipative noise in quadrature to $V_-$ leads to 
\begin{eqnarray}
 \xi^2_{\mathrm{TSS},\Gamma} & \approx & \frac{1 + \Gamma N t}{2N\beta}  + \frac{14}{9}\beta^2,  \label{eqn:TSSpert_superrad_crude}
\end{eqnarray}
where the prefactor $1/(2N\beta) \approx \mathrm{sin}^2(\psi)$ of the dissipative contribution corresponds to the degree which the excess noise is suppressed by the small squeezing angle when added in quadrature. In the limit 
$(\Gamma/\chi)\tau \ll S\tau^2$ Eqs.~(\ref{eqn:TSSpert_superrad_crude}) and (\ref{eqn:TSSpert_superrad}) become identical, indicating that this crude model captures the essential physics of TSS in the presence of 
collective emission.

\section{Single-particle dissipation \label{sec:SingleParticleDiss}}

In the main text we present a series of approximate analytic results pertaining to the squeezing which may be experimentally achieved in the presence of single particle decoherence. 
Specifically, we focus on the impact of single particle emission type processes (describing, e.g., spontaneous emission or Raman scattering) and dephasing 
(describing, e.g., Rayleigh scattering or dephasing due to background collisions and stray fields), with single particle jump operators $\hat{L}^{s}_j = \sqrt{\gamma_s/2}\hat{\sigma}^-_j$ 
and $\hat{L}^{el}_j = \sqrt{\gamma_{el}/8}\hat{\sigma}^z_j$ respectively. In this section we present the derivation of these results in more detail. 

The expressions we derive are crucially based on the fact that the relative strength of the elastic [$\chi = 4g^2\Delta_c/(4\Delta_c^2 + \kappa)$] and dissipative [$\Gamma = 4g^2\kappa/(4\Delta_c^2 + \kappa)$] 
interactions are controllable via the detuning $\Delta_c$ of the cavity field from the resonance with the atomic transition. To simplify the expressions in the following we assume that the cavity is operated  
in the limit $\Delta_c \gg \kappa$ ($\chi \gg \Gamma$) and thus we simplify $\chi \approx g^2/\Delta_c$ and $\Gamma \approx g^2\kappa/\Delta_c^2$. 


\subsection{One-axis twisting}

\subsubsection{Spontaneous emission}
To gain insight into the effects of single particle emission, $\hat{L}^{s}_j = \sqrt{\gamma_s/2}\hat{\sigma}^-_j$, on OAT we treat it as an independent (i.e., at the single-particle level) perturbative process on top of the squeezing.
This result is presented in Hu~\textit{et.~al.} Ref.~\cite{Vuletic2017}, though we include it here for completeness. In more detail, the emission is considered as a binomial random 
process that destroys the coherence between atoms in the ensemble. For an initial coherent spin state along $S_x$, after a duration $t$ an average of $\delta N \approx N(1-e^{-\gamma_s t})/2$ 
atoms are transferred from the single-particle state $\vert \uparrow \rangle$ to $\vert \downarrow \rangle$. This consequently leads to an increase in the variance of the spin distribution 
$\langle (\delta \hat{S}_z)^2 \rangle \equiv \langle (\delta \hat{S}_z )^2 \rangle_{sq} + Ne^{-\gamma_s t}(1 - e^{-\gamma_s t})/2$, where the subscript indicates the quadrature variance obtained 
from solely squeezing dynamics.

In the case of OAT, incorporating the effects of both single-particle and collective decoherence into Eq.~(\ref{eqn:OATpert}) in a similar manner to Eq.~(\ref{eqn:OATGamma_pert}) 
leads to the approximate expression 
\begin{equation}
 \xi^2_{OAT,\gamma_s} \approx \frac{1}{2N\beta} + N\Gamma t + 2\gamma_s t . \label{eqn:OATgammas_pert}
\end{equation}
Here, we have ignored the term $\propto \beta^2$ which describes the effects of the curved Bloch sphere as small in comparison to $\gamma_s t$ and $N\Gamma t$. 

The contribution due to collective emission can in principle be removed by detuning the cavity sufficiently far from resonance, $\Gamma/\chi = \kappa/\Delta_c \to 0$ for $\Delta_c \gg \kappa$. However, this 
would lead to a decrease in magnitude of $\chi\propto 1/\Delta_c$ and thus to a long time-scale for the one-axis twisting dynamics. Under this condition the system becomes highly vulnerable to other single particle noise sources, 
specifically spontaneous emission in this example, that would limit any achievable squeezing. To account for both collective and single-particle dissipation on an equal footing, one can instead 
optimise Eq.~(\ref{eqn:OATgammas_pert}) with respect to both the 
optimal squeezing time and the cavity detuning, instead of keeping a fixed $\Gamma/\chi$ as in the discussion of the previous sections. The cavity detuning is indeed a readily tunable parameter in the experiment.

Equation (\ref{eqn:OATgammas_pert}) is then minimized with respect to the squeezing duration $t_{\mathrm{opt}}$ and cavity detuning $\Delta_c^{\mathrm{opt}}$, to obtain
\begin{eqnarray}
 \left. \xi^2_{\mathrm{OAT}} \right\vert_{\gamma_s} \approx 6\left(N\eta_s\right)^{-1/3} , \label{eqn:OATgammas_bound}
\end{eqnarray}
for
\begin{equation}
 t_{\mathrm{opt}} = \gamma_s\left(\eta_s N \right)^{-1/3} , \quad 
 \Delta_c^{\mathrm{opt}} = \pm \frac{\kappa}{2} \sqrt{\frac{\eta_s N}{2}} ,
\end{equation}
where we define $\eta_s = 4g^2/(\kappa \gamma_s) \equiv \eta(\gamma/\gamma_s)$ as an effective single-atom co-operativity. From the optimal parameters we identify that 
$(N\Gamma t)_{\mathrm{opt}} = (2\gamma_s t)_{\mathrm{opt}}$, i.e., the cavity detuning is adjusted to minimize the collectively enhanced contribution $N\Gamma$ until it is of the same order 
as single-particle dissipation $\gamma_s$.

\subsubsection{Single-particle dephasing}

For the case of single-particle dephasing, $\hat{L}^{el}_j = \sqrt{\gamma_{el}/8}\hat{\sigma}^z_j$, our treatment is based on the exact solution of OAT dynamics in the presence of single-particle decoherence outlined in Refs.~\cite{FossFeig2013,FossFeig2013NJP}. 

First, we obtain the transformed correlation functions
\begin{eqnarray}
 \mathcal{A}_{el}(t) = e^{-\gamma_{el}t} \mathcal{A}(t)  , \notag  \\
 \mathcal{B}_{el}(t) = e^{-\gamma_{el}t/2} \mathcal{B}(t), \notag  \\
 \mathcal{C}_{el}(t) = e^{-\gamma_{el}t}\mathcal{C}(t) + (1 - e^{-\gamma_{el}t})S . \notag 
\end{eqnarray}
Moreover, if we ignore decay due to the shearing dynamics (valid for $N \gg 1$) we find that the effective spin length is $\vert|\langle \vec{S} \rangle\vert| = \vert|\langle \hat{S}_x(0) \rangle\vert|e^{-\gamma_{el}t}$. 

If we restrict dissipation to single-particle dephasing momentarily, the perturbative expression for the squeezing parameter can then be obtained similar to previous sections. 
Specifically, to appropriate order in $\beta \ll 1$ we find 
\begin{eqnarray}
 V_+V_- \approx \frac{S^2}{4} + \frac{2}{3}e^{-\gamma_{el}t}S^3\beta^3 , \\
 V_+ \approx 2S^2\beta e^{-\gamma_{el}t} .
\end{eqnarray}
Including the dissipative decay of the spin length, the squeezing parameter is then
\begin{eqnarray}
 \xi^2_{\mathrm{OAT},\gamma_{el}} \approx \frac{e^{2\gamma_{el} t}}{2N\beta} + \frac{2}{3}e^{\gamma_{el} t} \beta^2 . \label{eqn:OATgammael_only}
\end{eqnarray}

As previously, collective emission is incorporated as an independent perturbation to this expression Eq.~(\ref{eqn:OATgammael_only}), which leads to
\begin{equation}
 \xi^2_{\mathrm{OAT},\gamma_{el}} \approx \frac{e^{2\gamma_{el}t}}{2N\beta} + e^{\gamma_{el}t}N\Gamma t , \label{eqn:OATgammael_pert}
\end{equation}
where we remind the reader that we have assumed $\Gamma N t \ll 1$. We ignore the contribution $\propto \beta^2$ in this case as the squeezing is predominantly limited by dissipative noise 
much earlier than non-Gaussian effects. For $\gamma_{el}t \ll 1$ the perturbative expression can be furthered simplified to 
\begin{equation}
 \xi^2_{\mathrm{OAT},\gamma_{el}} \approx \frac{1 + 2\gamma_{el}t}{2N\beta} + \Gamma N t . 
\end{equation}
The suppression of dephasing in the final expression by the term $\propto 2N\beta$ illuminates that single-particle dephasing is far less detrimental to the squeezing when contrasted with collective or even single-particle emission. This is readily understood by noting 
that the dephasing commutes with the Hamiltonian. 

The best achievable squeezing is again obtained by optimising the cavity detuning to minimize the collective term $\Gamma N t$, whilst balancing against the excess noise due to dephasing at long times. Rigorously optimising 
Eq.~(\ref{eqn:OATgammael_pert}) in this manner is difficult. However, we can gain useful insight by assuming that the squeezing time is essentially insensitive to the dephasing and so follows the result for solely 
collective emission [i.e., Eq.~(\ref{eqn:OATGamma_bound}) and surrounding text]. We have checked this approximation with numerical calculations and find that it is reasonable for the regime of 
$\gamma_{el} t \lesssim 1$ considered. 

This approximation imposes that the dependence of the squeezing time on the cavity detuning is 
$t_{\mathrm{opt}} \approx [2\Delta_c^4/(g^6\kappa N^3)]^{1/3}$. It remains that we must optimize the squeezing with respect to the detuning $\Delta_c$, equivalent to solving 
\begin{eqnarray}
 0 & = & \frac{d}{d\Delta_c^{2/3}}\left[ \frac{e^{2\gamma_{el}C\Delta_c^{4/3}}}{\left(g^2 N C \Delta_c^{1/3}\right)^{2}} + e^{\gamma_{el}C\Delta_c^{4/3}}\frac{Ng^2\kappa C}{\Delta_c^{2/3}} \right] \notag \\
 & = & e^{C\gamma_{el}\Delta_c^{4/3}}\left(e^{C\gamma_el\Delta_c^{4/3}} + 2 \right)\left(2C\gamma_{el}\Delta_c^{4/3} -1 \right) ,
\end{eqnarray}
where we defined $C = [2/(g^6\kappa N^3)]^{1/3}$ for brevity. For $C\gamma_{el} \Delta_c^{4/3} \ll 1$ (which is checked from the subsequent solution for $\Delta_c$) we expand the exponentials to 
$\mathcal{O}(C\gamma_el\Delta_c^{4/3})$ to solve for the optimal detuning, thus finally obtaining
\begin{equation}
 \left. \xi^2_{\mathrm{OAT}}\right\vert_{\gamma_{el}} \approx \sqrt{\frac{4(41 + 13\sqrt{10})}{3\eta_{el}N}} \approx \sqrt{\frac{110}{\eta_{el}N}} , \label{eqn:OATgammael_bound}
\end{equation}
for 
\begin{eqnarray}
 t_{\mathrm{opt}} & = & \frac{\sqrt{10}-1}{6\gamma_{el}} , \\
 \Delta_c^{\mathrm{opt}} & = & \frac{\kappa}{2}\left[  \frac{\left(\sqrt{10}-1\right)\eta N}{12} \right]^{3/4} .
\end{eqnarray}
Here, the optimal squeezing time is fixed and weakly satisfies the requirement $\gamma_{el} t \lesssim 1$.

\subsection{Two-spin squeezing}

\subsubsection{Spontaneous emission}
Our treatment of TSS with single-particle spontaneous emission is based again on a semi-classical treatment of the dynamics, using the truncated Wigner approximation. Here, we consider the dynamics in the 
frame of the back-to-back spins but approximate the Hamiltonian to the simpler form $\hat{H} \approx \chi \hat{S}_y^2$. Whilst dissipation will inevitably introduce fluctuations in $\hat{S}_x$ and degrade the 
validity of this assumption, it will capture the essential physics for weak emission $\gamma_s t \ll 1$. 

Whilst the spontaneous emission is a single-particle process, we can approximately take it into account at the level of the collective spin variables. Specifically, we take the Lindblad form of the Heisenberg 
equations for single-particle emission. Summing over the single-particle equations and taking the mean-field approximation (see Sec.~\ref{sec:TSS_Ideal}) yields collective corrections to the TSS dynamics, 
resulting in the relevant equations:
\begin{eqnarray}
 \frac{dS_x}{dt} & = & 2\chi S_y S_z - \frac{\gamma_s}{2}S_x , \\
 \frac{dS_z}{dt} & = & -2\chi S_y S_x - \frac{\gamma_s}{2}\left( S_z + \frac{N}{2} \right) , \\
 \frac{dS_y}{dt} & = & -\frac{\gamma_s}{2}S_y , \label{eqn:TWA_Sy_emit} \\
 \frac{d\Delta_x}{dt} & = & 2\chi S_y\Delta_z -\frac{\gamma_s}{2}\Delta_x , \\
 \frac{d\Delta_z}{dt} & = & -2\chi S_y \Delta_x - \frac{\gamma_s}{2}\Delta_z .
\end{eqnarray}
Here, the difference variables are defined as previous, $\Delta_{\alpha} = S^{\alpha}_1 - S^{\alpha}_2$. We proceed according to the TWA method by solving the equations of motion for each variable and 
evaluating the relevant stochastic averages analytically. 

We proceed by first independently solving Eq.~(\ref{eqn:TWA_Sy_emit}) for $S_y(t) \equiv S_y(0)e^{-\gamma_s t/2}$. Solution of the remaining system of equations is simplified by assuming that for weak decoherence $\gamma_s t \ll 1$ 
we can substitute $S_y(t) \approx S_y(0)$ in the equations of motion. This approximation can be checked by comparing the final perturbative expression for squeezing to numerical solution of the 
full equations, for which we find good agreement. Essentially this approximation ignores secondary corrections to the shearing dynamics by dissipative decay of $S_y$. 
The remaining equations can then be solved exactly to obtain the relevant correlations:
\begin{eqnarray}
 \langle (\delta \Delta_z)^2 \rangle & = & \frac{S}{4}e^{-\gamma_s t} e^{-4S\chi^2t^2} \notag  \\
 & & \times \left[ 1 - 2S + (1+2S)e^{4S\chi^2t^2} \right] , \\
 \langle S_y \Delta_z \rangle  & = &  -S^2\chi t e^{-\gamma_s t} e^{-S\chi^2 t^2} , \notag \\
 \langle (\delta S_y)^2 \rangle  & = &  \frac{S}{2} e^{-\gamma_s t} .
\end{eqnarray}
The last expression for $\langle (\delta S_y)^2 \rangle$ is inconsistent physically and is an artefact of the approximate treatement of single-particle decoherence with our TWA approach. 
Physically, the variance along $\hat{y}$ is conserved throughout the evolution and must remain $\langle (\delta S_y)^2 \rangle = S/2$ if treated correctly at the single-particle level 
[i.e., it is a property that for Pauli matrices $(\hat{\sigma}^{\alpha}_i)^2 = 1$]. Hence we enforce strictly that $\langle (\delta S_y)^2 \rangle = S/2$ in the following calculation. 
The dependence on $\gamma$ of the remaining expressions $\langle (\delta \Delta_z)^2 \rangle$ and $\langle S_y \Delta_z \rangle$ essentially captures the effects of the dissipation at the single-body level, i.e., 
at the leading order dissipation can be treated independently from the many-body dynamics and simply added as an additional exponential decay of the correlations.

We proceed to calculate the perturbative squeezing expression as previously done, i.e., in terms of perturbative expansions for $\mathcal{A}$, $\mathcal{B}$, $\mathcal{C}$ 
for the small parameter $\beta = S\tau^2$ after substitution of the relevant correlation functions. 
Following this recipe, the squeezing parameter is obtained:
\begin{eqnarray}
 \xi^2_{\mathrm{TSS},\gamma_s} & \approx & \frac{1}{2N\beta} + \frac{2}{3}\beta^2 \notag \\
 & & + \gamma_s t \left( 1 - \frac{1}{2N\beta} - 2\beta + \frac{4}{3}\beta^2 \right) , \notag \\
 & \approx & \frac{1}{2N\beta} + \gamma_s t , \label{eqn:TSSgammas_only}
\end{eqnarray}
where the final expression encapsulates the effects of single-particle emission to lowest order. Here, whilst $\gamma_s t \ll 1$ we assume $\gamma_s t > \beta^2$, such that the decoherence is the 
limitation for achievable squeezing (i.e. added noise degrades squeezing before the non-Gaussian corrections to the distribution become relevant).

To obtain an expression for the squeezing taking into account both single-particle and collective emission, we note that the contribution of spontaneous emission is essentially at the single-body level in the above treatment. 
We thus assume that the correction $\gamma_s t$ can be similarly added to the squeezing expression Eq.~(\ref{eqn:TSSpert_superrad}) which takes into account collective decoherence. This leads to the approximate expression
\begin{eqnarray}
 \xi^2_{\mathrm{TSS},\Gamma,\gamma_s} \approx  \frac{1 + \Gamma N t}{2N\beta} + \gamma_s t . \label{eqn:TSSgammas_pert}
\end{eqnarray}
Optimisation of Eq.~(\ref{eqn:TSSgammas_pert}) with respect to both time and cavity detuning yields
\begin{eqnarray}
 \xi^2_{\mathrm{TSS}} \vert_{\gamma_s} \approx \sqrt{\frac{6\gamma_s\kappa}{g^2 N}} \equiv \sqrt{\frac{24}{N\eta_s}} ,
\end{eqnarray}
for 
\begin{eqnarray}
 t_{\mathrm{opt}} & = & \frac{2}{\gamma_s}\sqrt{\frac{2}{3\eta_s N}} , \\
 \Delta_c^{\mathrm{opt}} & = & \kappa \left( \frac{\eta_s N}{18} \right)^{1/4} ,
\end{eqnarray}
where we have defined the effective co-operativity $\eta_{s} = 4g^2/(\kappa \gamma_{s})$ as before. 

\subsubsection{Single-particle dephasing}
We treat TSS with single-particle dephasing using a semi-classical treatment similar to that presented above. As such, we simply present the resulting expressions here. 

The perturbative expression for the squeezing parameter, when we include only the effects of single-particle dephasing is given by 
\begin{eqnarray}
 \xi^2_{\mathrm{TSS},\gamma_{el}} \approx \frac{1}{2N\beta} + \gamma_{el} t , \label{eqn:TSSgammael_only}
\end{eqnarray}
for $\gamma_{el} t \ll 1$. 

For similar reasons, we can combine the result Eq.~(\ref{eqn:TSSgammael_only}) with collective emission in a manner identical to the previous subsection, resulting in 
the a final expression for squeezing:
\begin{eqnarray}
 \xi^2_{\mathrm{TSS},\Gamma,\gamma_{el}} \approx  \frac{1 + \Gamma N t}{2N\beta} + \gamma_{el} t . \label{eqn:TSSgammael_pert}
\end{eqnarray}
Subsequent optimisation of Eq.~(\ref{eqn:TSSgammael_pert}) with respect to both time and cavity detuning yields
\begin{eqnarray}
 \xi^2_{\mathrm{TSS}} \vert_{\gamma_{el}} \approx \sqrt{\frac{6\gamma_{el}\kappa}{g^2 N}} \equiv \sqrt{\frac{24}{N\eta_{el}}} ,
\end{eqnarray}
for 
\begin{eqnarray}
 t_{\mathrm{opt}} & = & \frac{2}{\gamma_{el}}\sqrt{\frac{2}{3\eta_{el} N}} , \\
 \Delta_c^{\mathrm{opt}} & = & \kappa \left( \frac{\eta_{el} N}{18} \right)^{1/4} , 
\end{eqnarray}
where we have again used the effective co-operativity $\eta_{el} = 4g^2/(\kappa \gamma_{el})$.

\begin{figure}
 \includegraphics[width=8cm]{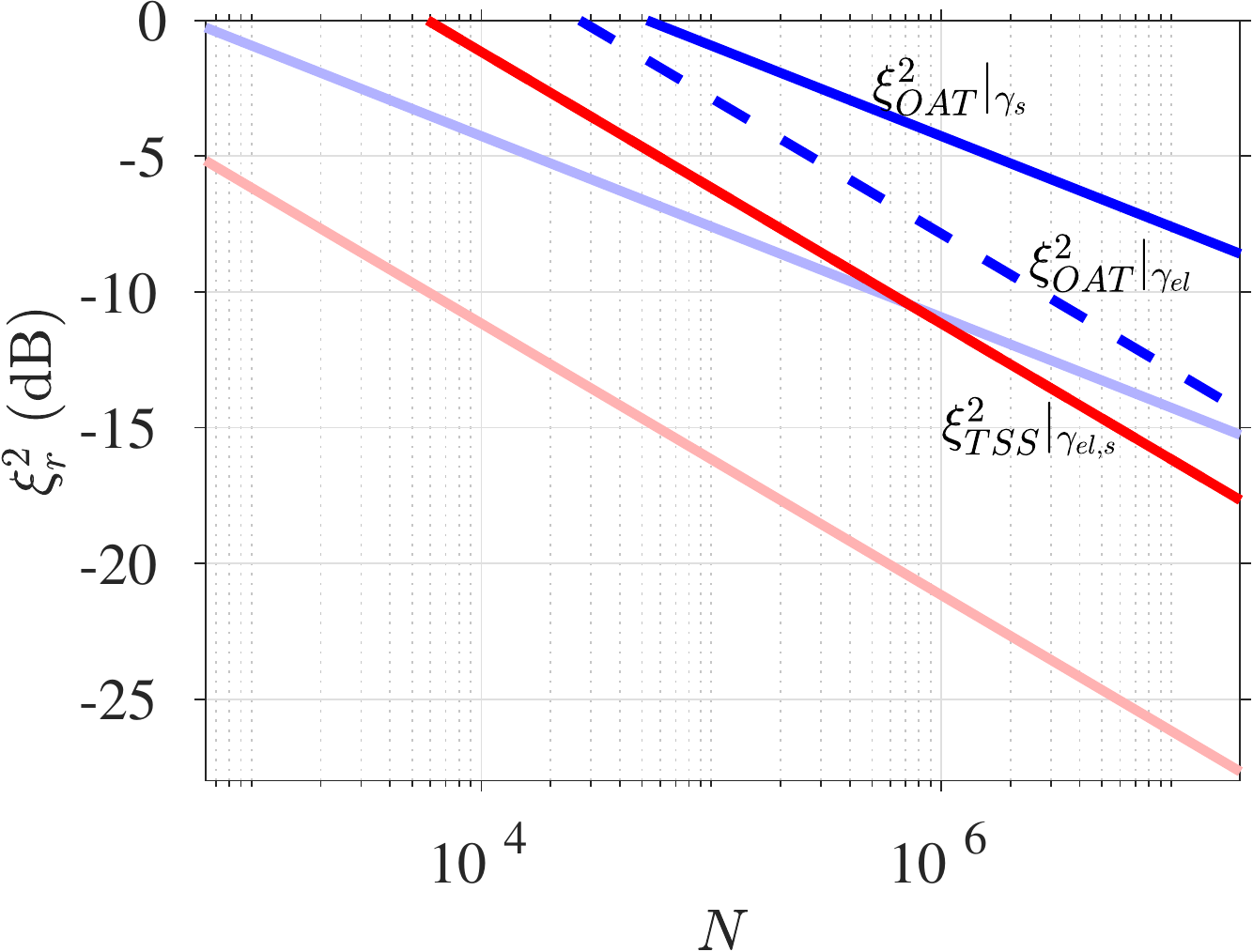}
 \caption{Comparison of optimized squeezing bounds for OAT (blue data) and TSS (red data). Cavity parameters for dark lines are: $\eta = 0.41$, 
 $\kappa/(2\pi) = 145$~kHz, $\gamma_s/(2\pi) = 0.1$~Hz and $\gamma_{el}/(2\pi) = 0.1$~Hz. Faded lines indicate the fundamental limits for OAT and TSS taking into account only the linewidth of the $^{87}$Sr 
 clock transition: $\eta = 0.41$, $\kappa/(2\pi) = 145$~kHz and $\gamma_s/(2\pi) \equiv \gamma/(2\pi) = 1$~mHz. }
 \label{fig:SqSingleParticle}
\end{figure}

For completeness, in Fig.~\ref{fig:SqSingleParticle} we compare the various analytic scaling relations for both OAT and TSS. Cavity parameters for dark lines are taken from state-of-the-art optical atomic clock and CQED experiments 
\cite{Norcia2017,Marti2018}: 
$\eta = 0.41$, $\kappa/(2\pi) = 145$~kHz, $\gamma_s/(2\pi) = 0.1$~Hz and $\gamma_{el}/(2\pi) = 0.1$~Hz. We also plot the fundamental bound on scaling when single-particle decoherence due to experimental 
imperfection can be ignored and thus squeezing is limited only by the linewidth of the $^{87}$Sr clock transition $\gamma_s/(2\pi) \equiv \gamma/(2\pi) = 1$~mHz. In the former case, we predict TSS can produce on the order of $\sim6$~dB  of squeezing 
below the standard quantum limit whilst OAT only produces $\sim1$~dB, both calculated with current state-of-the-art parameters and $N\sim10^5$. 
On the other hand, when the squeezing is limited only by fundamental noise sources we predict $\sim16$~dB for TSS compared to 
only $\sim8$~dB for OAT.

%
%

\section{Sensitivity of two-spin squeezing to number fluctuations}
Throughout the above sections we have assumed an idealized implementation of TSS. In particular, we have assumed that the initial state can be accurately prepared 
$\vert \psi_0 \rangle \equiv \vert N/4 \rangle_x \otimes \vert -N/4 \rangle_x$.
This assumption, that the initial state is an eigenstate of $\hat{S}_x$ with $\langle \hat{S}_x \rangle = 0$, is important as fluctuations in $\hat{S}_x$ can lead to a degradation of the achievable squeezing. 
In the ideal case,  the effect of such fluctuations is already observed as a minor decrease in the optimal squeezing compared to OAT [see Eq.~(\ref{eqn:TSSpert})]. 
An important question then is how robust the TSS protocol is to number fluctuations in each prepared atomic ensemble. 

To answer this question, we consider an initial state still composed of a pair of diametrically opposed collective spins, but introduce independent random fluctuations of zero mean and standard deviation $\sigma_n$ 
in the initial atom number of each ensemble (and thus fluctuations $\delta N \equiv \sqrt{2}\sigma_n$ in the total atom number). Such a state is characterized 
by the density matrix
\begin{widetext}
\begin{equation}
 \hat{\rho}_0 \equiv \int dn_1 dn_2 ~ P(n_1,n_2) \Big\vert \frac{N}{4} + \frac{n_1}{2} \Big\rangle \otimes \Big\vert -\frac{N}{4} + \frac{n_2}{2} \Big\rangle \Big\langle \frac{N}{4} + \frac{n_1}{2} \Big\vert \otimes \Big\langle -\frac{N}{4} + \frac{n_2}{2} \Big\vert , 
\end{equation}
\end{widetext}
where $P(n_1,n_2) = e^{-(n_1^2 + n_2^2)/2\sigma_n^2}/(2\pi\sigma_n^2)$.

\begin{figure}
 \includegraphics[width=8cm]{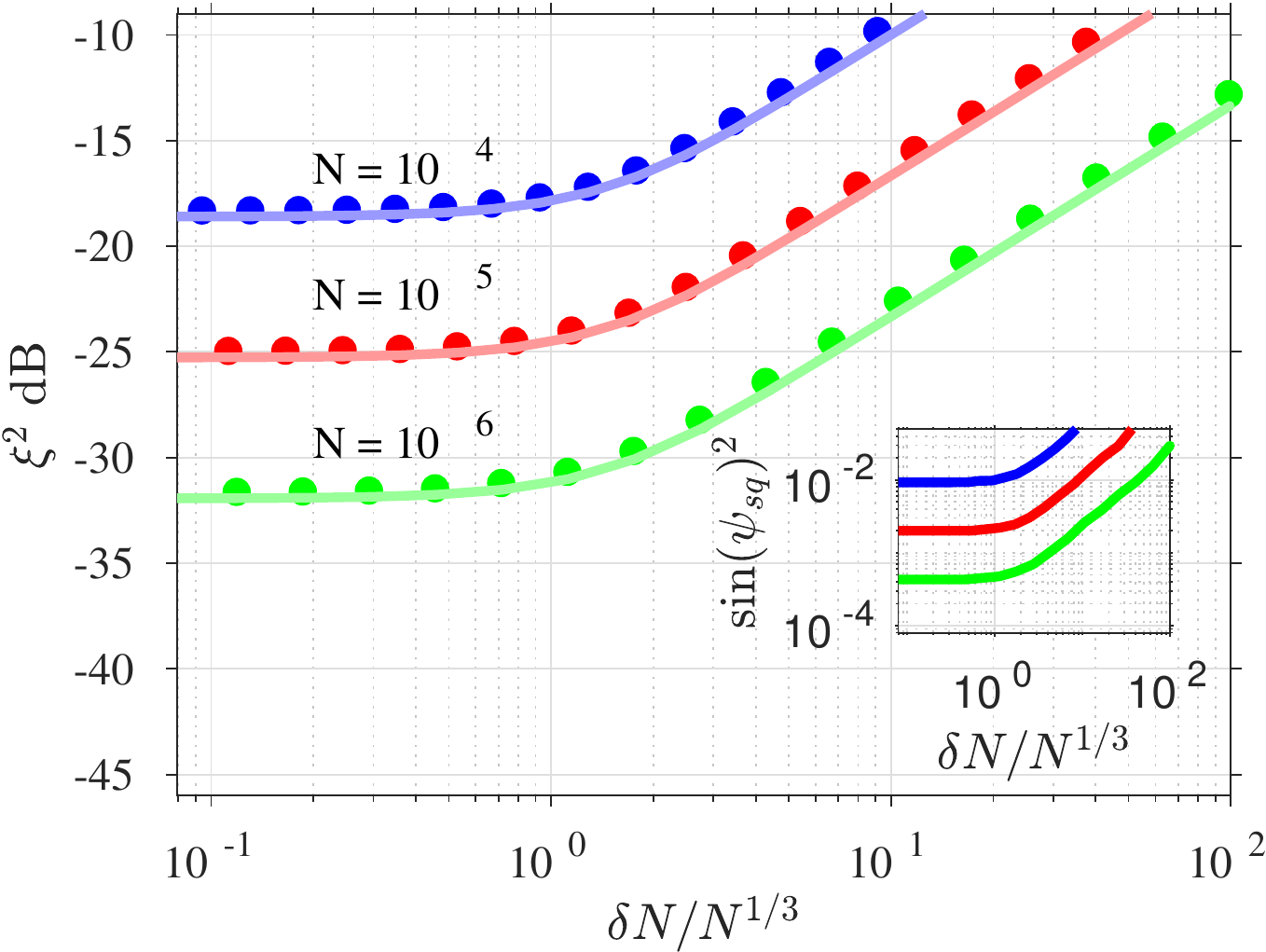}
 \caption{Effects of number fluctuations on squeezing generated with TSS. Markers indicate complete numerical calculation using TWA, whilst faded lines are numerical minimization of perturbative expression 
 Eq.~(\ref{eqn:TSS_NFluct}). We observe that ideal squeezing is preserved for $\delta N \lesssim N^{1/3}$. (Inset) Scaling of quantity $\mathrm{sin}^2(\psi_{sq})$ with number fluctuations. This quantity corresponds 
 to the prefactor suppressing dissipative noise due to collective emission in the treatment of TSS in Sec.~(\ref{sec:TSS_CollEmit}). We observe it remains at the ideal level for $\delta N \lesssim N^{1/3}$, indicating 
 the state remains heavily phase-squeezed.}
 \label{fig:NumFluct_TWA}
\end{figure}

The dynamics and squeezing of this initial state can again be solved using a semi-classical treatment. Essentially, one solves for the correlation functions as per the previous treatment using the TWA 
for arbitrary $n_1$ and $n_2$, and then performs a final averaging of the obtained expressions for the correlations with respect to the number fluctations, which can be evaluated analytically. 
A perturbative expression for the squeezing is then obtained, 
\begin{equation}
 \xi^2_{\mathrm{TSS},\sigma_n} \approx \frac{1}{2N\beta} + \frac{16\sigma_n^2}{N}\beta + \frac{14}{9}\beta^2 . \label{eqn:TSS_NFluct}
\end{equation}
The number fluctuations lead to non-Gaussian corrections emerging at lower order, $\propto \beta$, destroying squeezing much earlier. 
From inspection of Eq.~(\ref{eqn:TSS_NFluct}) and comparison of the terms, we argue that fluctuations become important when $\sigma_n \sim N^{1/3}$. 
The reasoning for this is as follows: In the absence of fluctuations, squeezing occurs on the time-scale $\tau \sim N^{-2/3}$, due to the emergence of the usual non-Gaussian corrections (oversqueezing) 
described by the term $\beta^2$ in Eq.~(\ref{eqn:TSS_NFluct}). We thus claim that fluctuations in the atom number will become important when $\sigma_n^2/N \sim \beta$, i.e. the latter two 
terms in Eq.~(\ref{eqn:TSS_NFluct}) become comparable, which leads to the estimate $\sigma_n \sim N^{1/3}$. For $\sigma_n \gtrsim N^{1/3}$ the term $\propto \beta^2$ can then be disregarded 
in Eq.~(\ref{eqn:TSS_NFluct}), and the optimal squeezing in the limit of `large fluctuations' will then be given by
\begin{equation}
 \xi^2_{\mathrm{TSS}}\vert_{\sigma_n} \approx \frac{\sqrt{2}\sigma_n}{N} = \frac{\delta N}{N}. \label{eqn:TSSbound_numfluct}
\end{equation}
where we introduce the fluctuation in total atom number $\delta N \equiv \sqrt{2}\sigma_n$.
We illustrate these arguments in Fig.~(\ref{fig:NumFluct_TWA}), using a full numerical TWA calculation, wherein we plot the squeezing parameter as a function of total fluctuation in 
atom number $\delta N$ for $N = 10^4$, $10^5$ and $10^6$. 

Having established that the squeezing generated by TSS is reasonably robust to number fluctuations, a final question is to investigate whether the protocols retains its robusticity to collective emission. 
From a qualitiative perspective, we conjecture that the protocol should remain robust for at least $\sigma_n \lesssim N^{1/3}$. This is based on the previous discussion, which showed that TSS remains 
unaffected until fluctuations reach this level. This implies that the state remains phase-squeezed, and thus should retain its robusticity to collective emission. We support this reasoning in the inset of
Fig.~\ref{fig:NumFluct_TWA}, by plotting the quantity $\mathrm{sin}^2(\psi_{sq})$, which corresponds to the trigonometric factor suppressing the collective emission in the crude treatment of dissipation 
in Sec.~\ref{sec:TSS_CollEmit}. We observe that this quantity remains approximately unchanged from the ideal ($\sigma_n = 0$) level for total number fluctuations $\delta N \lesssim N^{1/3}$, 
consistent with this reasoning. 

\begin{figure}
 \includegraphics[width=8cm]{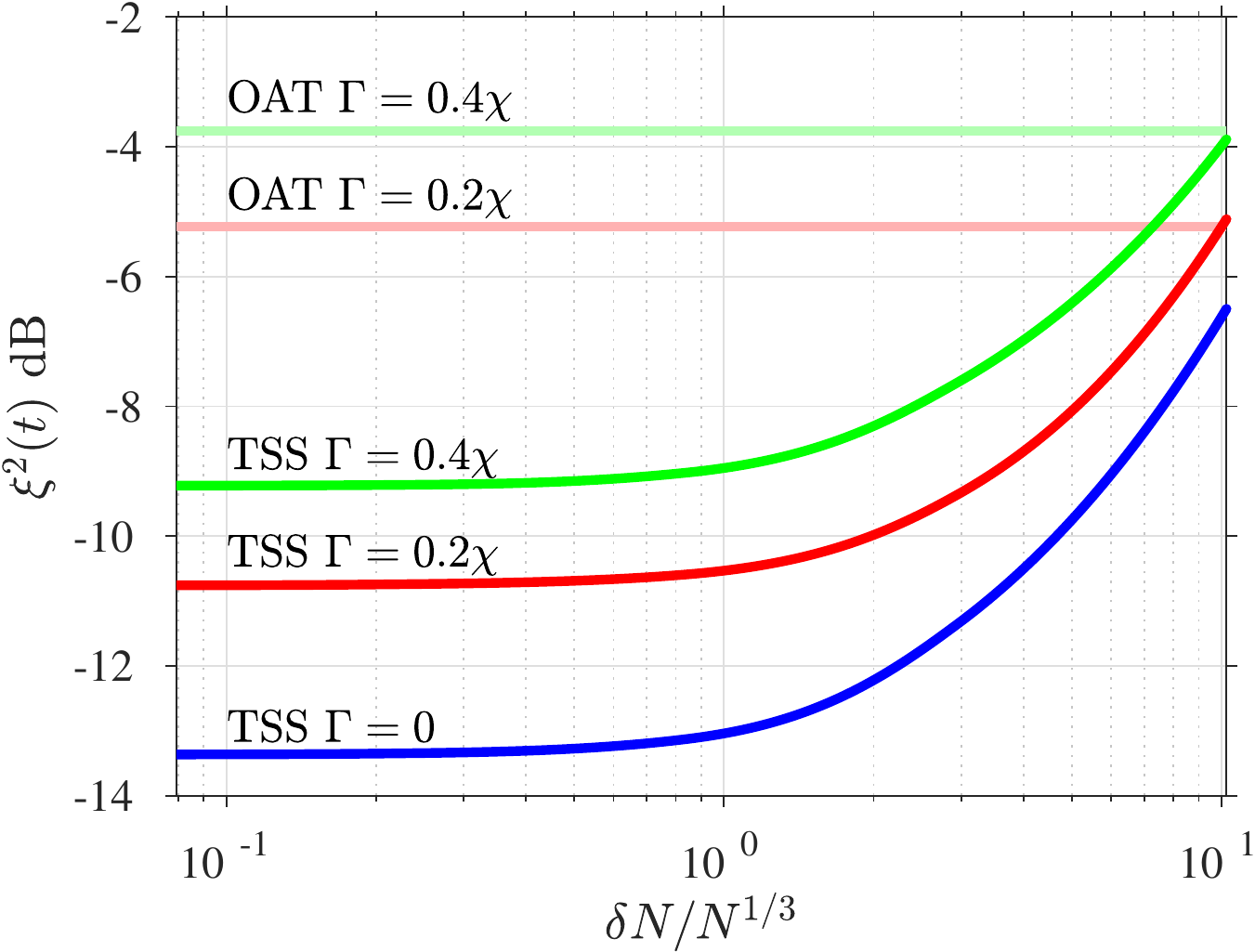}
 \includegraphics[width=8cm]{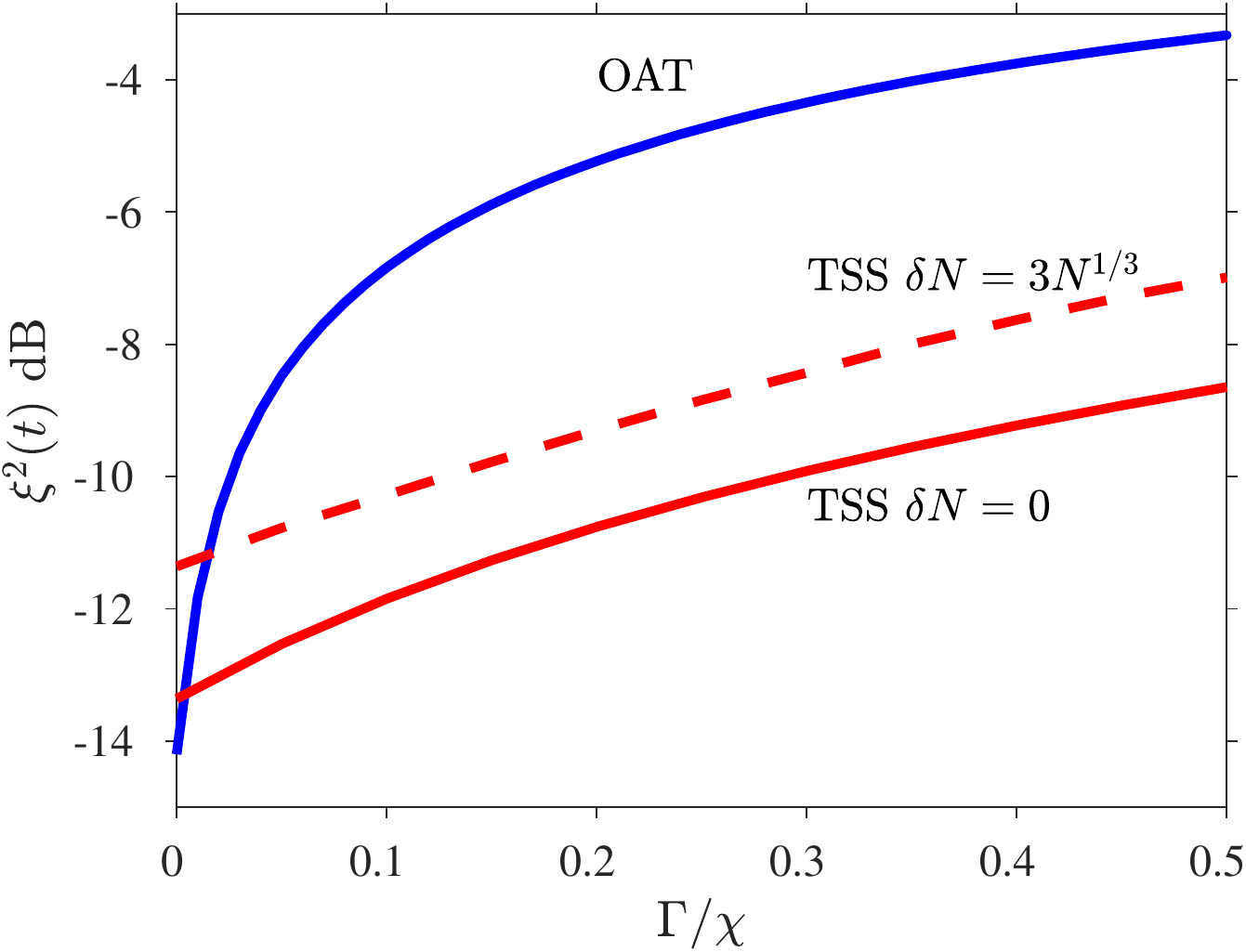}
 \caption{Effects of number fluctuations on robustness of TSS to collective emission. Results are for $N = 200$ using a (truncated) exact numerical calculation. 
 (Top) Scaling of squeezing with number fluctuatons $\delta N$ for fixed dissipation strength $\Gamma$. The TSS protocol remains at the ideal ($\delta N = 0$) level until a 
 crossover at $\delta N \sim N^{1/3}$. For this system size, TSS remains superior to OAT until $\delta N \gg N^{1/3}$, with the dominant limitation being the fundamental loss of squeezing generated by TSS rather than 
 increased sensitivity to emission. (Bottom) Scaling of squeezing with dissipation strength $\Gamma$. The robusticity and scaling of TSS remains close to the ideal $\delta N = 0$ level for reasonable 
 fluctuations, $\delta N \sim N^{1/3}$. We also compare to the scaling of OAT for clarity.}
 \label{fig:NumFluct_exact}
\end{figure}

To more rigorously justify this argument we use (truncated) exact calculations for a small system $N = 200$, as shown in Fig.~\ref{fig:NumFluct_exact}. We observe that  
TSS remains robust to collective emission for reasonable fluctuations $\delta N \lesssim N^{1/3}$. In particular, we observe no large qualitative change in scaling with $\Gamma$ as 
fluctuations are introduced, consistent with our conjecture and the results indicating the squeezing remains predominantly in the phase-quadrature in the inset of Fig.~\ref{fig:NumFluct_TWA}. 
As such, we argue that the primary barrier for TSS in the presence of number fluctuations is the ability to generate squeezing (see prior discussion) and 
that robusticity to collective emission is a secondary issue. 

\bibliography{TwoSpinSqueezing_refs}

\end{document}